\documentclass[10pt,a4paper]{article}
\usepackage{amsmath,amssymb,graphicx,bm}
\usepackage[utf8]{inputenc}
\usepackage{multicol,booktabs,lastpage}
\pdfoutput=1
\usepackage{subcaption,xcolor,braket,float}
\usepackage[top=1in,left=1in,right=1in,bottom=1in]{geometry}
\usepackage[autostyle,english=american]{csquotes}
\MakeOuterQuote{"}
\providecommand{\keyword}[1]{\textbf{Keywords:} #1}
\providecommand{\pacs}[1]{\textbf{PACS:} #1}

\title{Effects of Finite Size of Constituent Quarks on Nucleon-Nucleon Interaction.}
\author{Vanamali C. Shastry\footnote{vanamalishastry@gmail.com}\quad  K. B. Vijaya Kumar\footnote{kbvijayakumar57@gmail.com}\\
Department of Physics, Mangalore University, Mangalore, India.}
\begin{document}
\maketitle

\begin{abstract}
We have investigated the effect of the finite size of the constituent quarks on singlet and triplet nucleon-nucleon potentials, obtained in the framework of the $SU(2)$ nonrelativistic quark model using the resonating group method in the Born-Oppenheimer approximation. The full Hamiltonian used in the investigation includes the kinetic energy, two-body confinement potential, one gluon exchange potential (OGEP), and instanton induced interaction (III). The effects of the smearing of the contact interactions and the variation of the constituent mass of the quarks are discussed.
\end{abstract}

\keyword{Nucleon-nucleon interaction, nonrelativistic quark model, resonating group method, constituent quarks.}


\section{Introduction}

Quantum chromodynamics (QCD) is widely accepted as the fundamental theory of strong interactions. However, low energy QCD is not solvable using field theoretic techniques due to the nonperturbative nature of the coupling constant. Strongly interacting particles also exhibit confinement, the exact nature of which has not been computed from QCD. This creates the need for effective theories and models. \par

QCD based nonrelativistic quark models (NRQMs) and effective field theories (EFTs) have been used extensively in the past to study nucleon-nucleon (NN) interaction. EFTs have been successful in explaining the long range behavior of NN interaction without taking in to account the quark structure of nucleons \cite{Machleidt:1989tm,Weinberg:1991um,Entem:2003ft,Ordonez:1995rz,Epelbaum:1999dj}. A similar approach is used to explain the intermediate part using the exchange of $\sigma$ meson between nucleons. However, there are questions on the existence of $\sigma$ mesons (\cite{Agashe:2014kda} and references there in). In spite of these successes, EFTs neither delve into the structure of nucleons nor do they take in to account the effect of the quark structure of nucleons on NN interaction. This lacuna left by the EFTs is addressed by the quark models \cite{VijayaKumar:1993np,Myhrer:1987af,Maltman:1983wx,Neudatchin:1977vt,Faessler:1983qh,Zhang:1994pp,Stancu:1997dq,Breit:1960zz}.\par

According to quark models, hadrons are made up of constituent quarks. Spontaneous breaking of the chiral symmetry of the QCD modifies the quark propagator and provides an effective, momentum dependent mass to the quarks \cite{Diakonov:1995zi}. The effective mass of the constituent quarks at zero momentum is approximately $1/3^{rd}$ of the nucleon mass. NRQMs use potentials derived from QCD to model the interaction between constituent quarks. NRQMs have been employed successfully in the past to study NN interaction \cite{Myhrer:1987af,Shimizu:1989ye,Oka:1989ud}. These models indicate that the NN interaction at short distances should be governed by the QCD dynamics. Particularly, the studies using one gluon exchange potential (OGEP) suggest that the short range repulsion arises due to the spin-spin interaction between constituent quarks.

Oka et al. \cite{Oka:1989ud} have studied the NN interaction using the instanton induced interaction (III) instead of the one pion exchange potential (OPEP). The color magnetic term of the III provides short-range repulsion. Thus the short range repulsion of the NN interaction is attributed to the exchange part of the color magnetic interaction of both the OGEP and the III \cite{Oka:1989ud}. The justification for the inclusion of the III is given in \cite{Vanamali:2016qwr}. \par

In our earlier work \cite{Vanamali:2016qwr}, the resonating group method (RGM) technique in the framework of the NRQM was  employed to obtain the NN adiabatic potential for the singlet ($^1S_0$) and triplet ($^3S_1$) states using the Born-Oppenheimer approximation. The aim was to understand the role played by the OGEP, III, and OPEP to the adiabatic NN potential. The study indicated that the short-range repulsion arises from the kinetic energy and exchange terms of the color magnetic terms of the OGEP and III.\par

 Chu \textit{et. al.} \cite{Chu:1994vi} have studied the effects of the instantons on the structure of the hadrons by including the effects of the gluons and instantons using lattice QCD. This study concludes that instantons contribute significantly to the low energy structure of the hadrons. Instantons modify the quark propagator and give an effective mass to the quarks. This means the constituent quarks are nothing but bare quarks surrounded by a cloud of virtual particles which provides an effective size to quarks. The contact interaction between constituent quarks implies that the virtual clouds surrounding the two constituent quarks do not have any effect on the hyperfine splitting of the hadron spectrum. Hence, the study of the effects of the smearing of the quarks on NN interaction becomes essential.\par  
 Also, in the quark models of mesons, the $\pi-\rho$ splitting comes from the color magnetic term of OGEP which is a $\delta(r)$ potential and hence is singular. Since $\pi-\rho$ splitting is large, the color magnetic term of OGEP cannot be treated perturbatively, it is required to smear it, so as to incorporate the finite size of quark core. A nonperturbative potential model analysis of the light and heavy mesons by Stanley and Robson \cite{Stanley:1980zm} found that the rms radius of the constituent quarks must be of the order of $0.15\text{ fm}$ for the model to reproduce the rms radius of the pion. Though the specific form of the form factor is not crucial, it is necessary to use a finite range function, to avoid the collapse of an attractive $\delta(r)$ potential \cite{Bhaduri:1981pn}.\par
Further, in an NRQM, the size of the hadrons, characterized by their rms radii, is given by the oscillator size parameter. This picture is a naive one, as the only contribution to the size of the hadrons comes from the dominant nonperturbative part of the Hamiltonian. The contributions of the perturbative part of the Hamiltonian, which decide the rest of the properties of the hadron spectra, are completely neglected. This leads to an erroneous result for the rms radius of the proton at 0.5 fm. However, in the real life implementations of the NRQM, the oscillator size parameter is considered a free parameter. We can overcome this shortcoming of the model by assuming the constituent quarks to be of finite size. A previous approach suggested by Povh and Hufner, the constituent quarks were assumed to have a radius inversely proportional to their constituent mass \cite{Povh:1987ju,Povh:1990ad}.\par
In a study based on the NJL model, Lutz and Weise have argued that the low energy quark-antiquark polarization effects strongly screen the valence quarks (which are bare quarks) and hence lead to "spatially extended structures" \cite{Lutz:1990zc}. In their words, "constituent quarks are not point-like". The authors successfully reproduce the rms charge radius of the proton by redefining it as 
\[
	\langle r^2_E\rangle_{\text{proton}} = \langle r^2\rangle_{\text{core}} + \langle r^2\rangle_{\text{cloud}} 
\]
where $\langle r^2\rangle_{\text{cloud}}$ arises due to the quark-antiquark polarization effects.\par
 It is clear that the quark masses play an important role in the dynamics of the short range repulsion. Since no attempt have been made in the constituent quark models to study the effect of the quark masses on the short range part of the NN interaction, we have investigated the NN interaction by taking into effect the finite size of the constituent quarks by replacing the Dirac delta function by the Gaussian function. The full Hamiltonian used in the investigation includes the kinetic energy, two-body confinement potential, OGEP, and III, which includes the effect of quark exchange between the nucleons. The contribution of the OGEP and III to the NN adiabatic potential is discussed. The finite size effect of the potential on $^1S_0$ and $^3S_1$ NN potentials are obtained in the framework of the SU(2) NRQM using RGM in the Born-Oppenheimer approximation.\par
In the present work, we have modeled the constituent quarks as soft spheres with the bare quarks in the center surrounded by meson cloud. Thus, to simplify the calculations, we have incorporated the finite size effects into the potential by smearing the $\delta$-function.\par
The aim of the present investigation is to make a detailed study of the contribution of the color magnetic part of OGEP and III, by replacing the delta term in the color magnetic term of OGEP and in the direct and exchange part of the III potential by  the Gaussian function so as to take into account the finite size of the quarks and hence taking into account the chiral symmetry, the basic symmetry of the QCD.\par

The paper is organized as follows: the following section discusses the model used in the study, section 3 gives a brief review of the RGM, section 4 presents the results of the work and section 5 gives the summary of the work. \par

\newpage
\section{The Model} 
The Hamiltonian used in this study has the form,

\begin{eqnarray}
H & = & K + V_{int} + V_{Conf}-K_{CM}\\
&=& H_0+V_{int}\nonumber
\end{eqnarray}
where K is the kinetic energy, $V_{int}$ is the interaction potential term and $V_{Conf}$ is the harmonic confinement potential and $K_{CM}$ is the kinetic energy of the center of mass. The interaction potential is,
\begin{equation}
	 V_{int} = V_{OGEP} + V_{III}
\end{equation}
where,
\begin{equation}
	V_{OGEP}=\dfrac{\alpha_{s}}{4}\sum_{i<j}\left(\dfrac{1}{r_{ij}}-
	\dfrac{\pi}{m_{q}^{2}}(1+\dfrac{2}{3}\bm\sigma_{i}.\bm\sigma_{j})
	\delta(\bm r_{i}-\bm r_{j})\right)\bm\lambda_{i}.\bm\lambda_{j}
\end{equation}

\begin{equation}
	V_{Conf}=-a_{c}\sum_{i<j} r_{ij}^{2}\bm\lambda_{i}.\bm\lambda_{j}
\end{equation}

In the $SU(2)$ limit, the III potential takes the form \cite{Vanamali:2016qwr}, 
\begin{equation}\label{eq:iii}
	V_{III}=-\dfrac{1}{2}W\sum_{i<j}\left(\dfrac{16}{15}+\dfrac{2}{5}\bm\lambda_{i}.
\bm\lambda_{j}+\dfrac{1}{10}\bm\sigma_{i}.\bm\sigma_{j}\bm\lambda_{i}.\bm\lambda_{j}\right)
\delta(\bm r_{i}-\bm r_{j})
\end{equation}

In the above expressions, $r_{ij}$ is the separation between the quarks, $m_{q}$ is the mass of the quark, $\bm\sigma_{i}$ is the spin of the $i^{th}$ quark, W is the strength of III potential and $a_{c}$ is the confinement strength parameter.

 In the previous work, we had shown that the above Hamiltonian reproduces the qualitative features of NN interaction potential \cite{Vanamali:2016qwr}. The primary inferences from the earlier study and their relevance to the present work are listed below.

\begin{enumerate}
	\item The short range repulsion in the NN interaction arises due to color magnetic forces and the kinetic energy of the quarks. The orbital component of the color magnetic terms of both OGEP and III are given by delta functions. Hence, smearing of delta function will affect the short-range repulsion significantly. However, the kinetic energy of the quarks remains unchanged.
	\item The color singlet component of the III is attractive in the short range. This behavior will also be affected by the smearing of the delta function.
	\item The contributions of the confinement potential and the OPEP remain the same. Single pion exchange between quarks is known to contribute slightly to the repulsive core \cite{Vanamali:2016qwr} and since we are not interested in the effects of the finite size of pions, we omit the OPEP in the present study.
\end{enumerate}

To study the effects of the finite size of constituent quarks, we replace $\delta(| \bm r_i - \bm r_j|)$ by $\dfrac{1}{\pi^{3/2}r_0^3} e^{-r_{ij}^2/r_0^2}$, where $r_0$ represents the size of the quarks. A discussion on the effect of the smearing is in order. Delta function potential represents contact interactions which are zero range interactions. A juxtaposition of the contact interactions with the constituent quark picture leads to inconsistencies. A constituent quark acquires dynamical mass due to the chiral symmetry breaking. The dynamical mass and hence the constituent quark can be modeled as a virtual cloud surrounding the bare quark. Assuming contact interactions between constituent quark would discount all the interactions between the virtual clouds surrounding the bare quarks as well as between the virtual cloud and the opposing bare quark. This negates the very idea of the dynamical mass of the constituent quarks. Hence, it is absolutely necessary that a smeared potential be used. \par
The smearing of the delta function leads to two things:
\begin{enumerate}
	\item The strength of the quark-quark interactions increases gradually as quarks approach each other as opposed to an abrupt interaction on contact. Thus the constituent quarks "see" more and more of each other's virtual cloud thereby experiencing a force that increases gradually.
	\item Smearing also results in a finite range for color magnetic interactions. The smearing parameter ($r_0$) decides the relative strength at a distance $r$ away from the bare quark. Thus, we can interpret $r_0$ as the effective size of the constituent quark.
\end{enumerate}
Smearing gives rise to a picture where constituent quarks are modeled as soft spheres instead of point particles. \par
The Yukawa smearing, which has been used to study meson spectra, is known to reproduce the light meson and baryon spectra if the constituent quark size parameter is chosen to be around $0.4 \text{ fm} - 0.5 \text{ fm}$ \cite{Bhaduri:1981pn}. The Yukawa smearing has been used to study light baryon resonances \cite{Vijande:2003gk}. The Gaussian approximation of delta function has been used to study heavy baryon spectra \cite{SilvestreBrac:1996bg} as well as quarkonium spectra \cite{Ono:1982ft}. \par

\section{Resonating Group Method}
Here, the RGM is employed to calculate the NN interaction potential \cite{Oka:1981ri}. The trial wave function used to calculate the Hamiltonian matrix elements is the ground state wave function of harmonic oscillator potential given by the following equation where $A$ and $B$ represent the two quark clusters (nucleons). 

\begin{eqnarray}
\nonumber \phi(\bm r_A) = \dfrac{1}{(\pi b^2)^{9/4}} \prod_{i=1}^{3} exp(-\dfrac{1}{2b^2}(\bm r_i - \dfrac{\bm s_I}{2})^2)\\
\phi(\bm r_B) = \dfrac{1}{(\pi b^2)^{9/4}} \prod_{i=4}^{6} exp(-\dfrac{1}{2b^2}(\bm r_i + \dfrac{\bm s_I}{2})^2)
\end{eqnarray}

The following equation is solved to find the NN potential.
\begin{equation}\label{eigeq}
	\bra\psi (H-E)\mathcal{A} \ket\psi = 0
\end{equation}

where, $\mathcal{A} =\dfrac{1}{10}(1-9P_{36}^{OSTC})$ is the anti-symmetrization operator and $P_{36}^{OSTC}$ is the quark exchange operator that exchanges the orbital ($O$), spin ($S$), isospin ($T$) and color ($C$) quantum numbers of quarks 3 and 6. The energy ($E$) obtained is projected to the $l=0 $ channel. \par

\section{Results and Discussion}
The parameters in the present model are: $\alpha_s\,,\, b\,,\, m_q\,,\, r_0\,,\, W\,,\, \text{and}\,  a_c$. The strong coupling constant $\alpha_s$ and the III strength parameter $W$ are chosen to reproduce the $N-\Delta$ mass splitting. The confinement strength parameter, $a_c$ is fixed from the stability condition $\dfrac{\partial H_0}{\partial b} = 0$, where $H_0$ is the unperturbed Hamiltonian. The oscillator size parameter ($b$) is also a free variable in our model. The value of $b$ is unchanged from the previous work \cite{Vanamali:2016qwr}. \par

The expressions for $\alpha_s$ and $W$ are given below.

\begin{eqnarray}
	\alpha_s&=&C_1 m_q^2(r_0^2+b^2)^{3/2}(M_\Delta - M_N)_{OGEP}\\
	W&=&C_2(r_0^2+b^2)^{3/2}(M_\Delta - M_N)_{III}
\end{eqnarray}
where, $(M_\Delta - M_N)_{OGEP,III}$ are the contributions of OGEP and III to the $N-\Delta$ mass difference and $C_{1,2}$ are numerical constants. The stability condition results in the following expression for $a_c$,
\begin{equation}
	a_c = \dfrac{C_3}{m_q b^4}
\end{equation}
where, $C_3$ is a constant.\par

One should note that the finite size of the constituent quarks affects the short range $\delta$-interactions much more than the long range Coulomb forces. The smearing of the $\delta$-interactions influences the strong coupling constant $\alpha_s$ and hence the Coulomb interactions.

\subsection{Smearing of the contact interactions}
We begin by analyzing the effect of variation of $r_0$ on various components of III and color magnetic part of OGEP. The confinement potential and the kinetic energy of the quarks are independent of $r_0$. \par

\begin{figure}[h]
	\centering
	\begin{subfigure}[b]{0.3\textwidth}
		\includegraphics[scale=0.2]{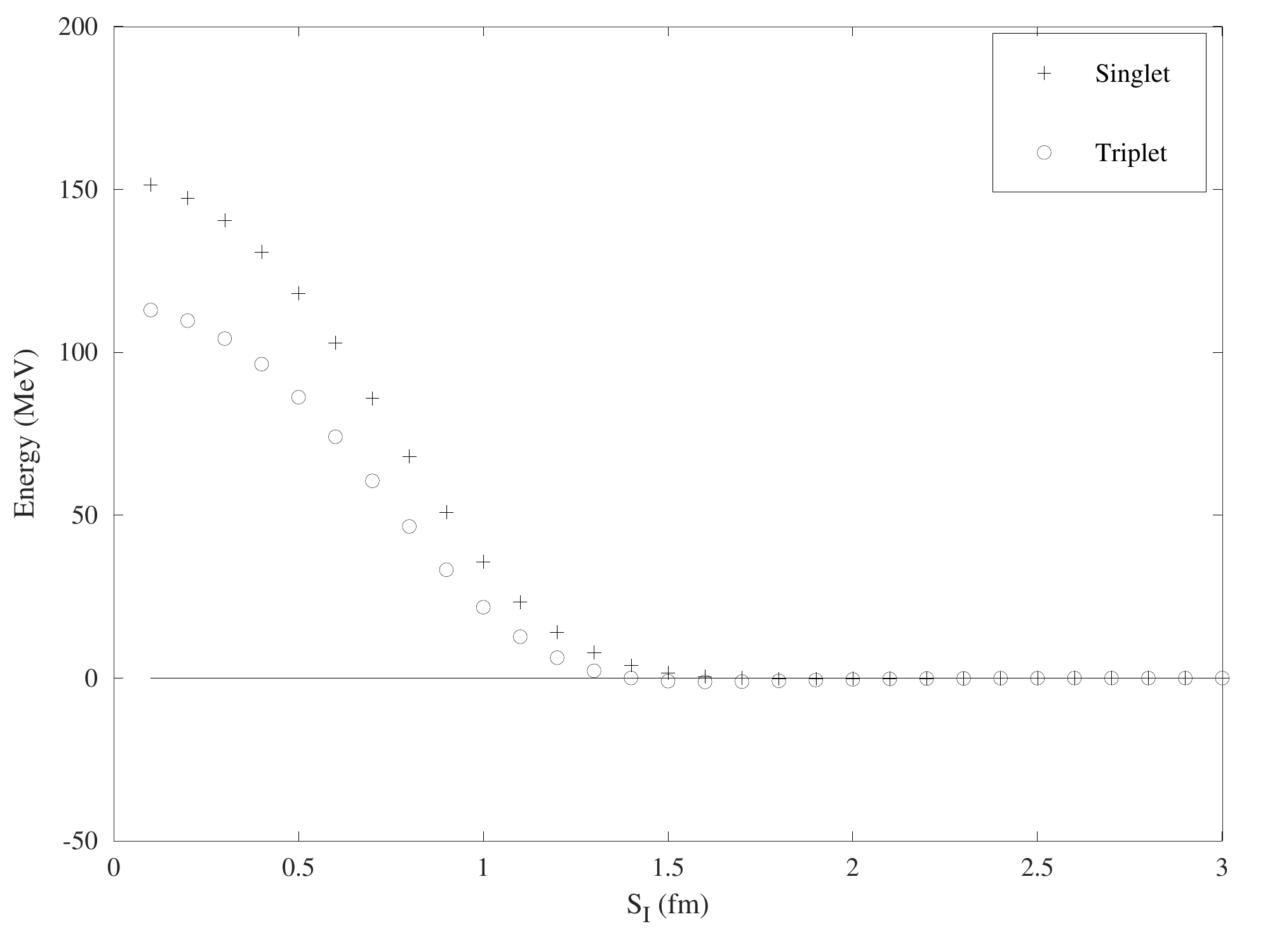}
		\caption{$r_0=0.1\text{ fm}$}
	\end{subfigure}
	\begin{subfigure}[b]{0.3\textwidth}
		\includegraphics[scale=0.2]{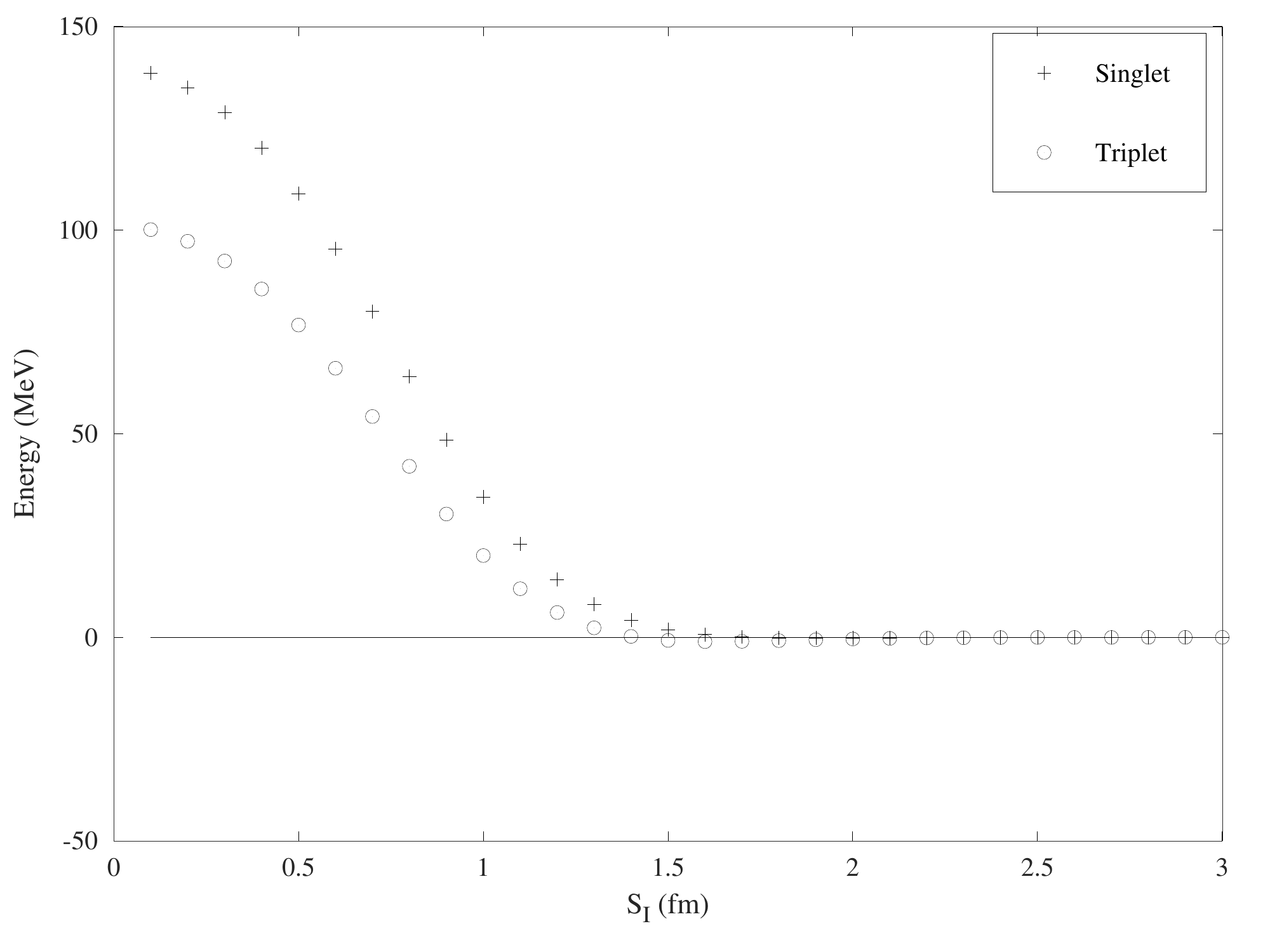}
		\caption{$r_0=0.2\text{ fm}$}
	\end{subfigure}
	\centering
	\begin{subfigure}[b]{0.3\textwidth}
		\includegraphics[scale=0.2]{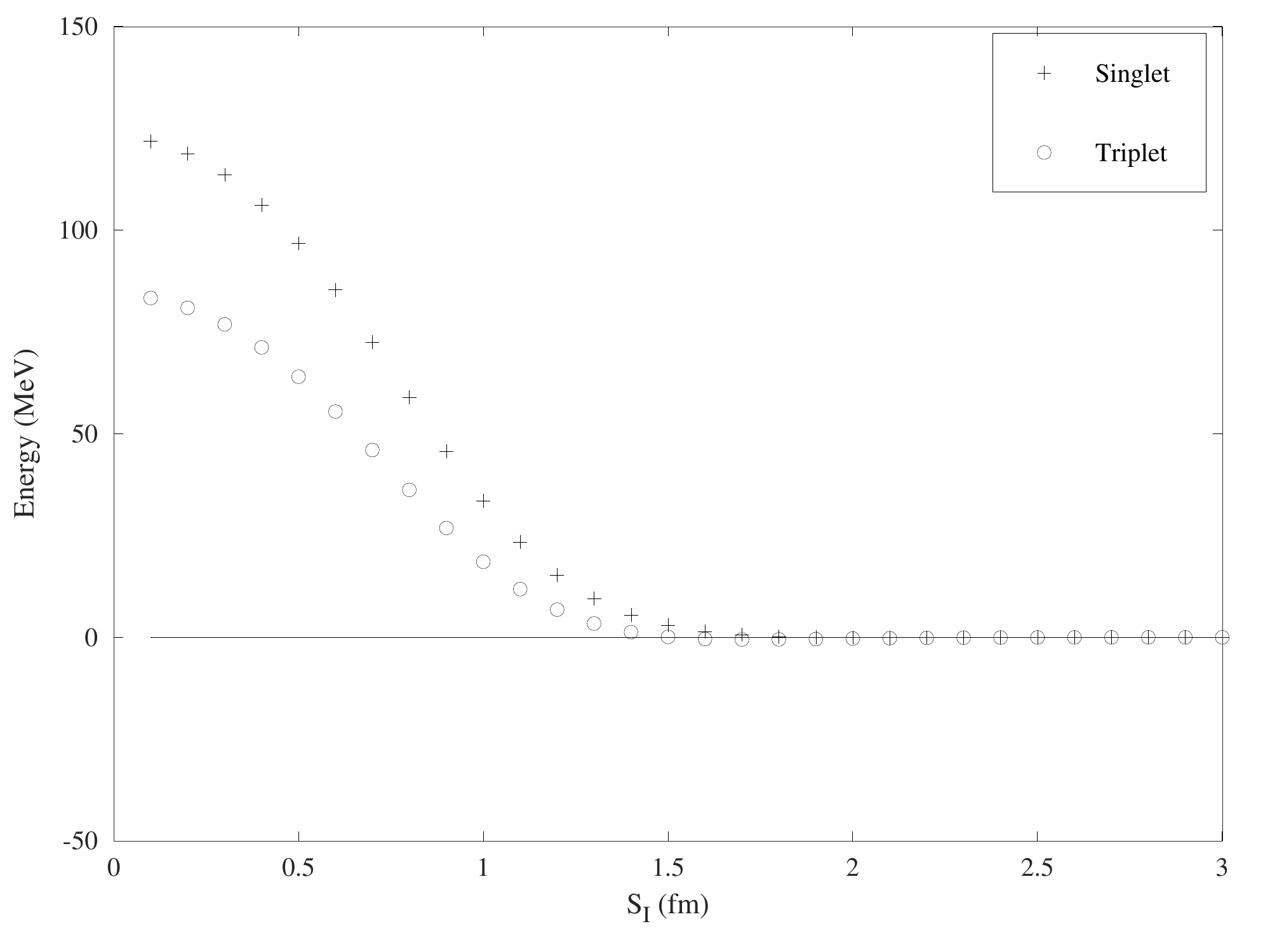}
		\caption{$r_0=0.3\text{ fm}$}
	\end{subfigure}
	\centering
	\begin{subfigure}[b]{0.3\textwidth}
		\includegraphics[scale=0.2]{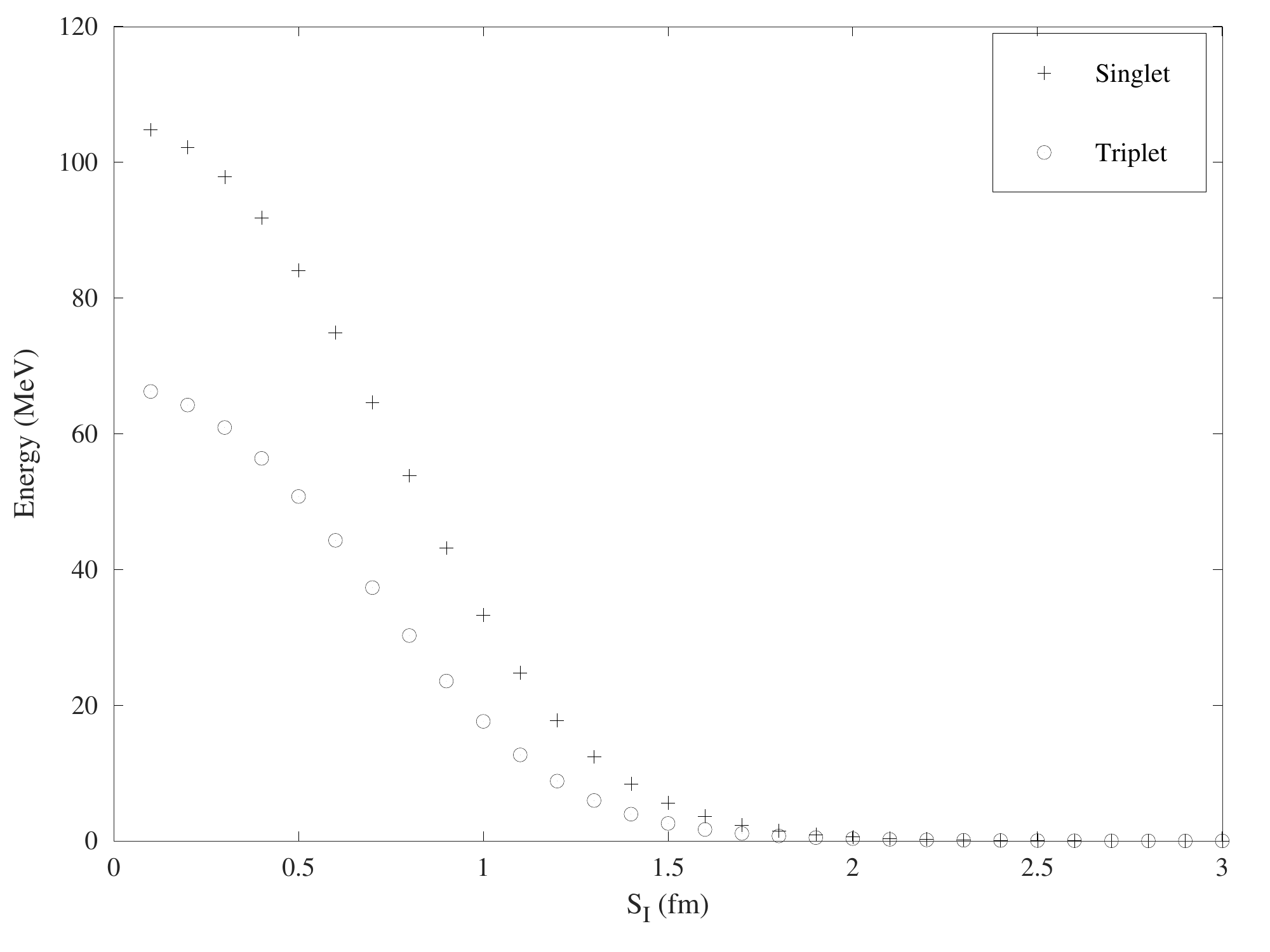}
		\caption{$r_0=0.4\text{ fm}$}
	\end{subfigure}
	\centering
	\begin{subfigure}[b]{0.3\textwidth}
		\includegraphics[scale=0.2]{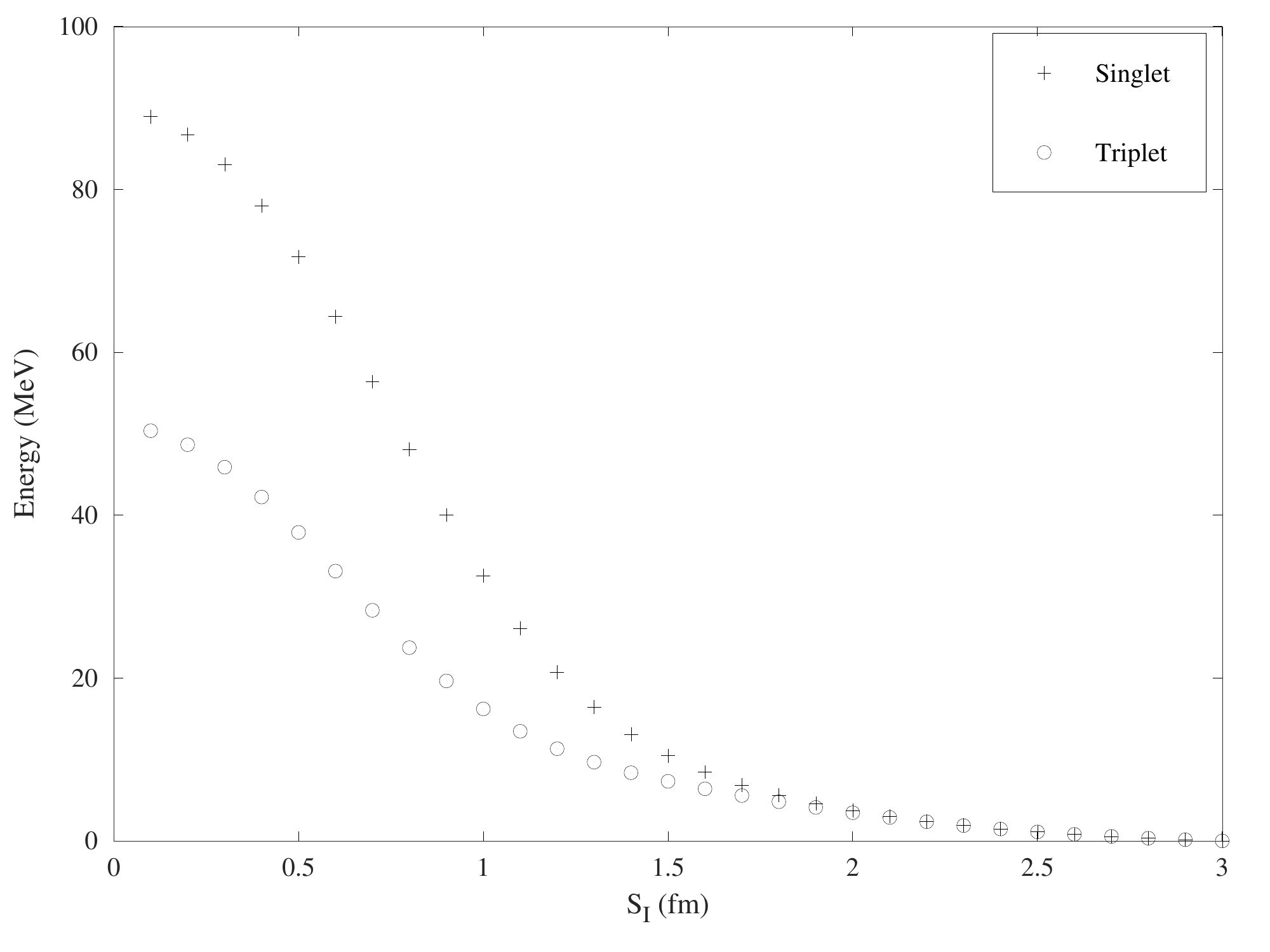}
		\caption{$r_0=0.5\text{ fm}$}
	\end{subfigure}
	\centering
	\begin{subfigure}[b]{0.3\textwidth}
		\includegraphics[scale=0.2]{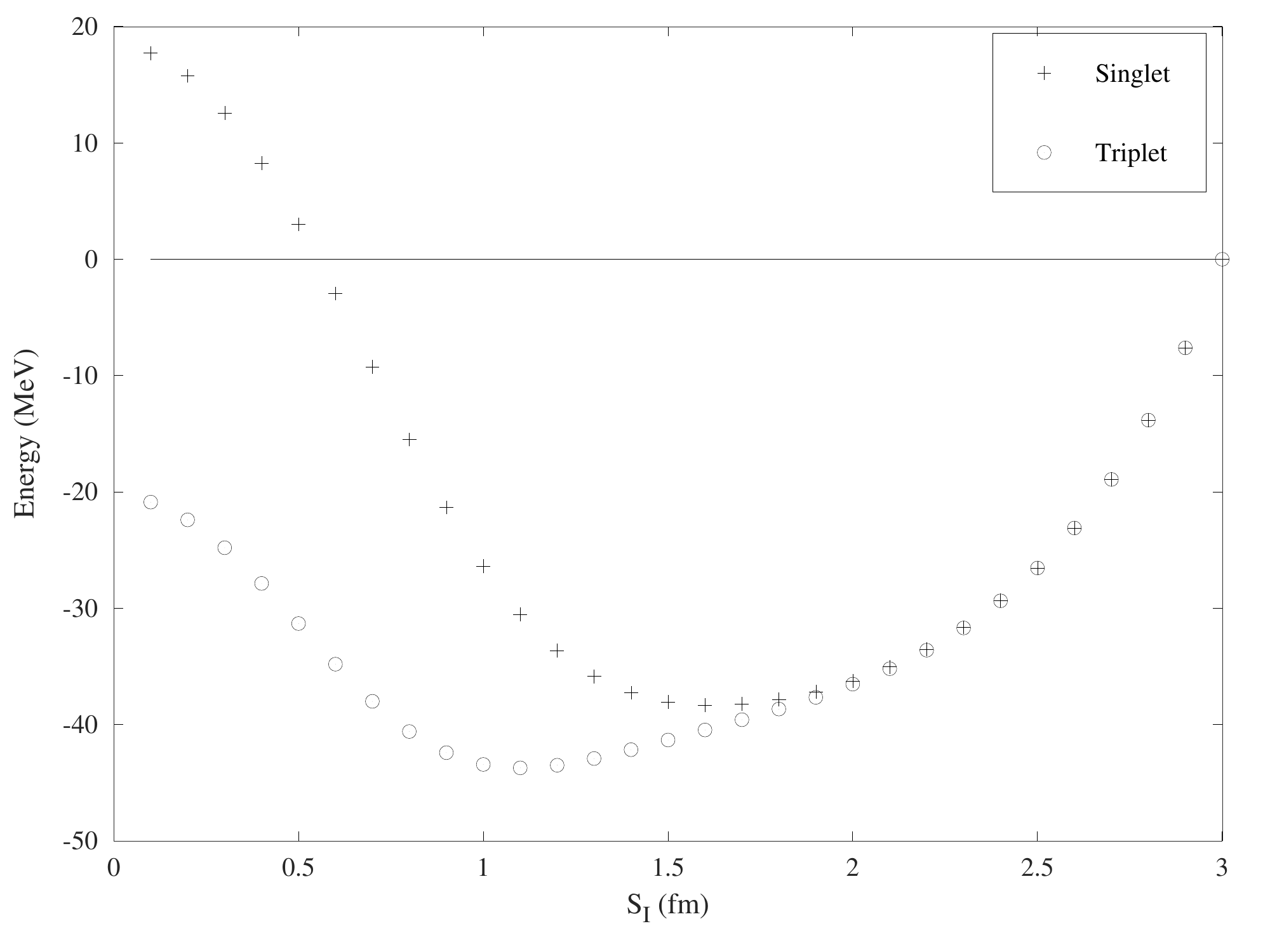}
		\caption{$r_0=0.6\text{ fm}$}
	\end{subfigure}
	\caption{Variation of total adiabatic NN potential ($m_q=300\text{ MeV}$) in MeV against the internucleon separation $S_I$ in fm with respect to $r_0$. In the plots, "o" are for $^1S_0$ potential and "$+$" are for $^3S_1$ potential. The horizontal line corresponds to 0 MeV.}\label{rep2}
\end{figure}

The magnitude of short range repulsion reduces as $r_0$ increases. Also, the range of short range interactions reduces as $r_0$ increases. When $r_0$ is very closer to b, the NN potential loses the intermediate range attraction due to the dominant color magnetic interactions of OGEP and eventually becomes completely attractive. The NN potentials for different values of $r_0$ are shown in fig. (\ref{rep2}). The qualitative features of the NN adiabatic potential do not change for $r_0<0.4\text{ fm}$ . The exchange part of the Hamiltonian is repulsive and the $^1S_0$ state is more repulsive than $^3S_1$ state. The attractive nature of III is consistent with the earlier calculations. The strength of attraction of III remains constant, but the range of III increases significantly when $r_0 > 0.4\text{ fm}$. The color magnetic part of OGEP is repulsive for $r_0<b$. \par

\begin{figure}[t]
	\centering
	\begin{subfigure}[b]{0.3\textwidth}
		\includegraphics[scale=0.2]{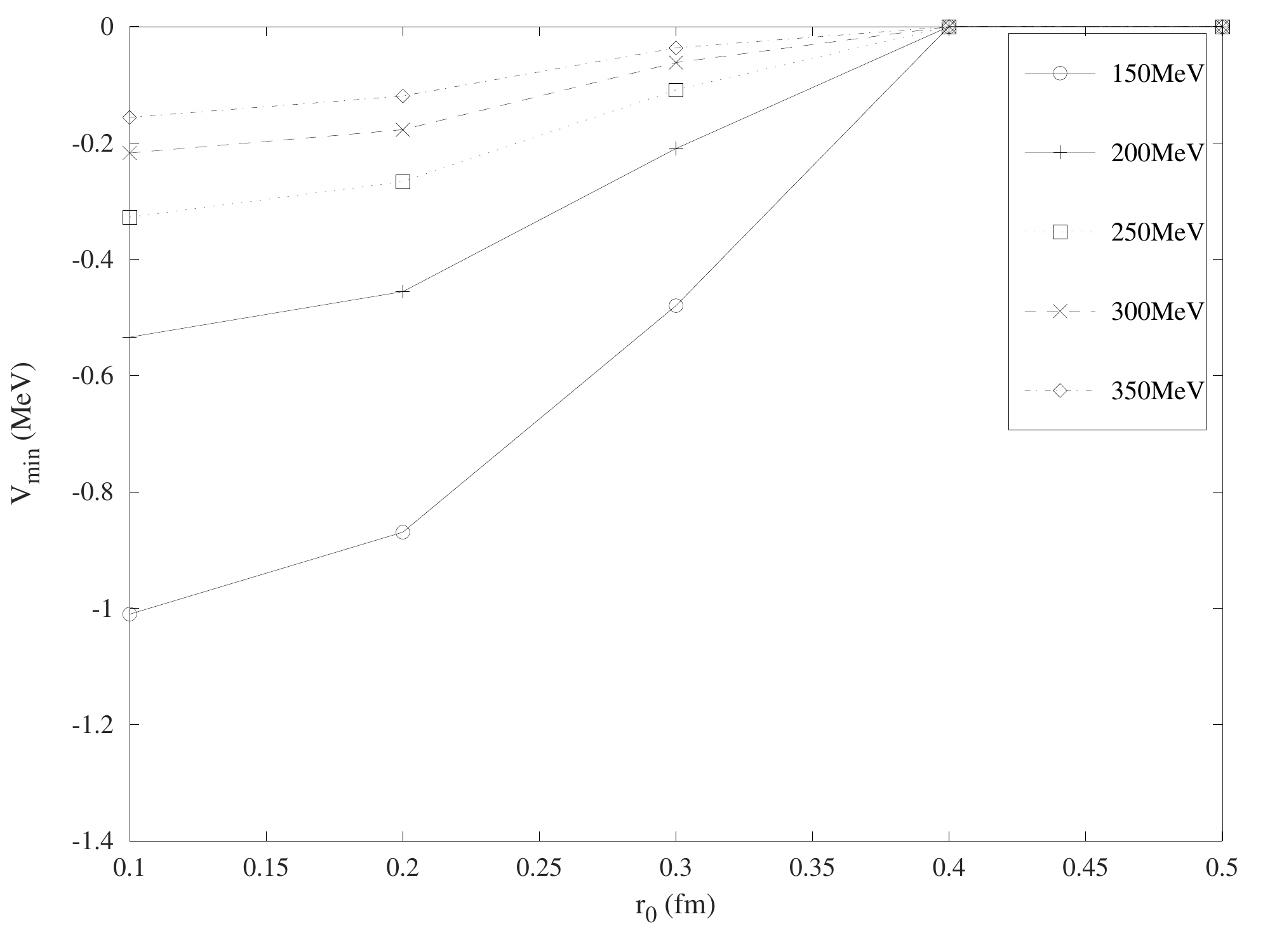}
	\end{subfigure}
	\centering
	\begin{subfigure}[b]{0.3\textwidth}
		\includegraphics[scale=0.2]{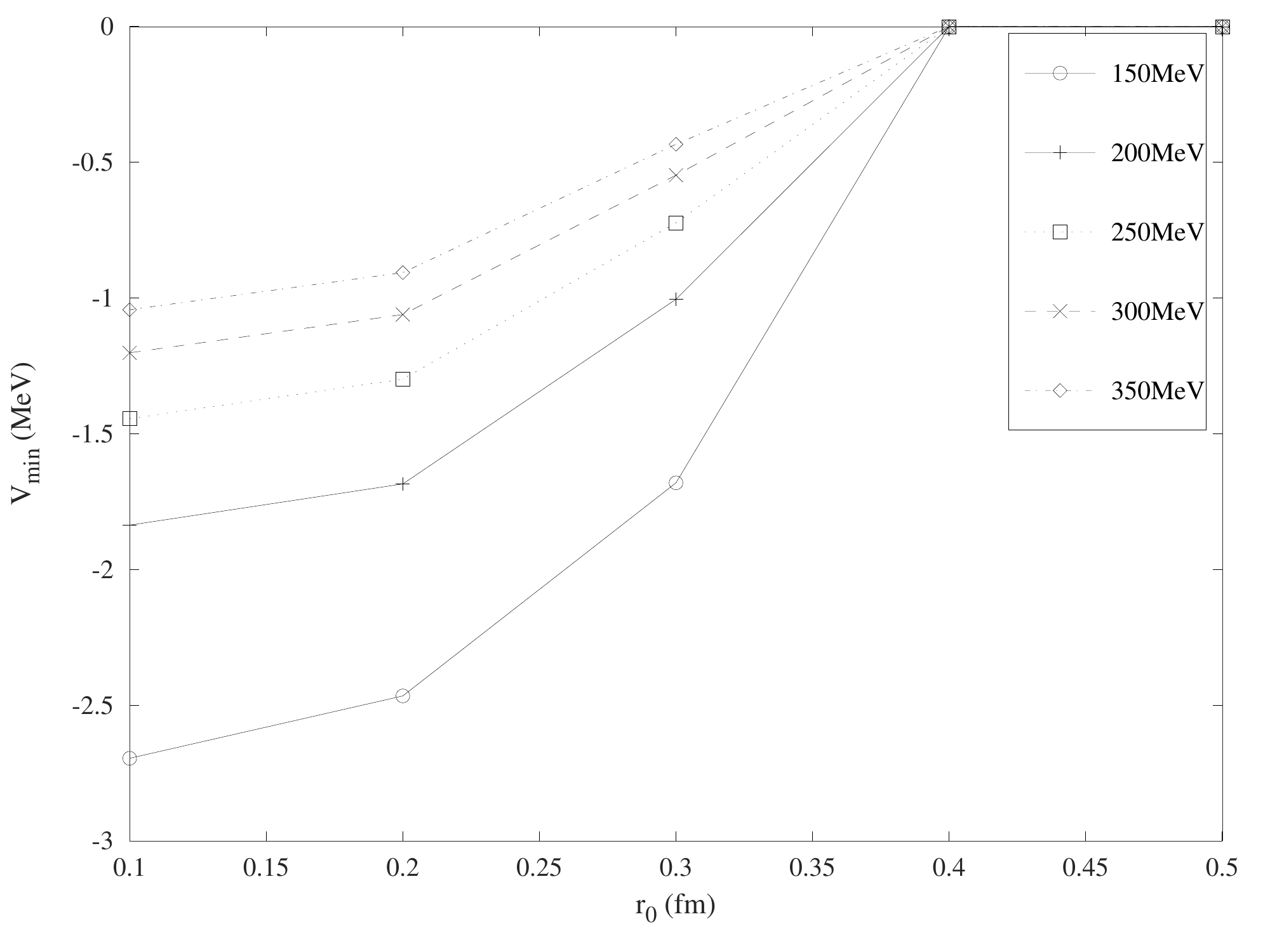}
	\end{subfigure}
	\centering
	\begin{subfigure}[b]{0.3\textwidth}
		\includegraphics[scale=0.2]{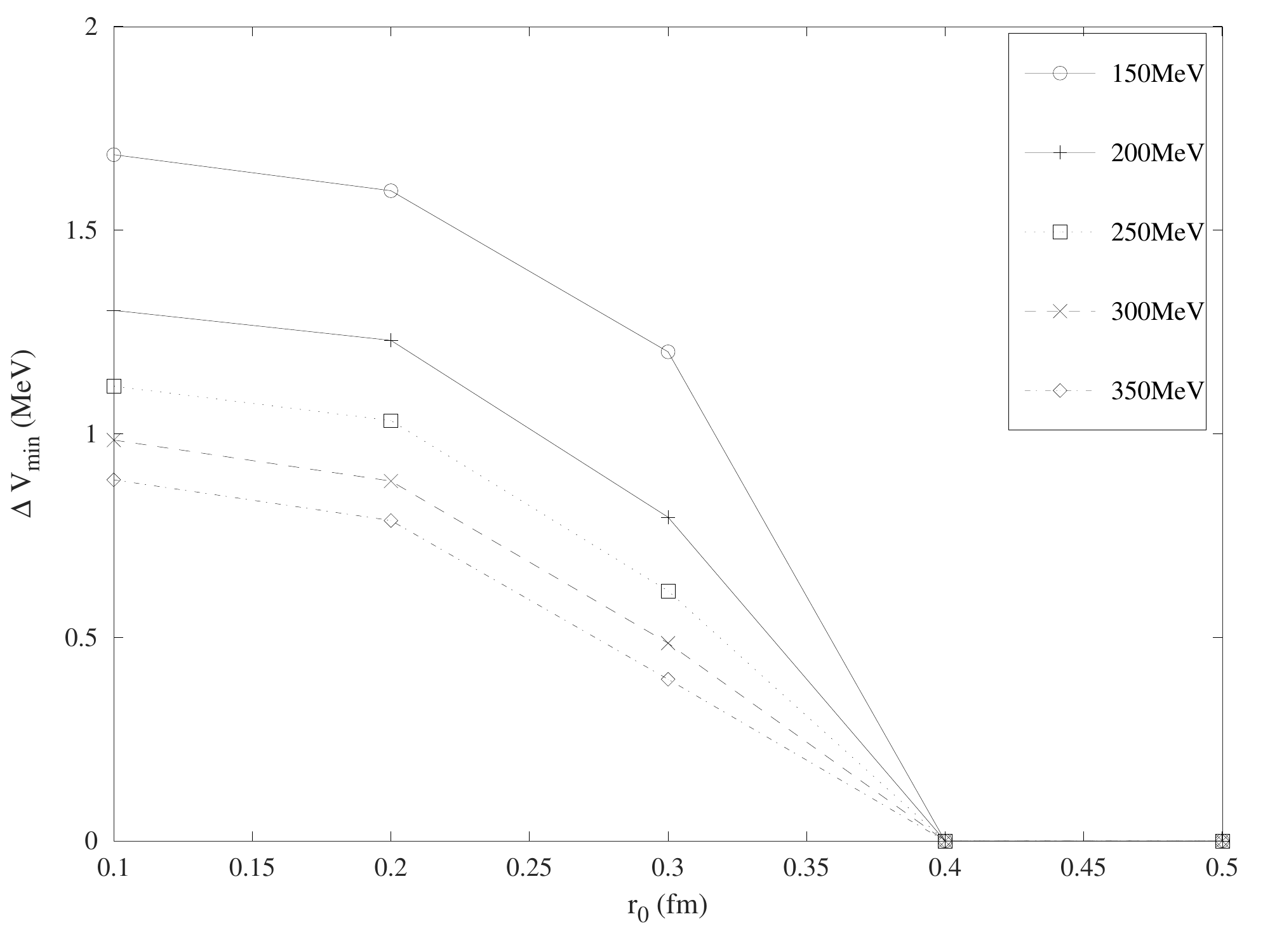}
	\end{subfigure}
	\caption{Variation of the minimum of the NN potential in the $^1S_0$ (left) and $^3S_1$(center) channel and the difference between the minima of the two channels (right) as functions of $r_0$.}\label{vmin}
\end{figure}

This indicates that the repulsion arises due to contact interactions between constituent quarks. The strength of repulsion of the color magnetic part of OGEP decreases as $r_0$ increases. Also, the effective range of color magnetic part of the OGEP increases significantly for $r_0>0.4\text{ fm}$. The variation of the minimum of the NN potential as a function of quark size parameter is shown in fig. (\ref{vmin}). The NN potential has the largest attraction when $r_0$ is the closest to $b$. The attraction reduces in magnitude as $r_0$ reduces and saturates as $r_0$ approaches zero. \par

\begin{figure}[H]
		\centering
	\begin{subfigure}[b]{0.3\textwidth}
		\includegraphics[scale=0.2]{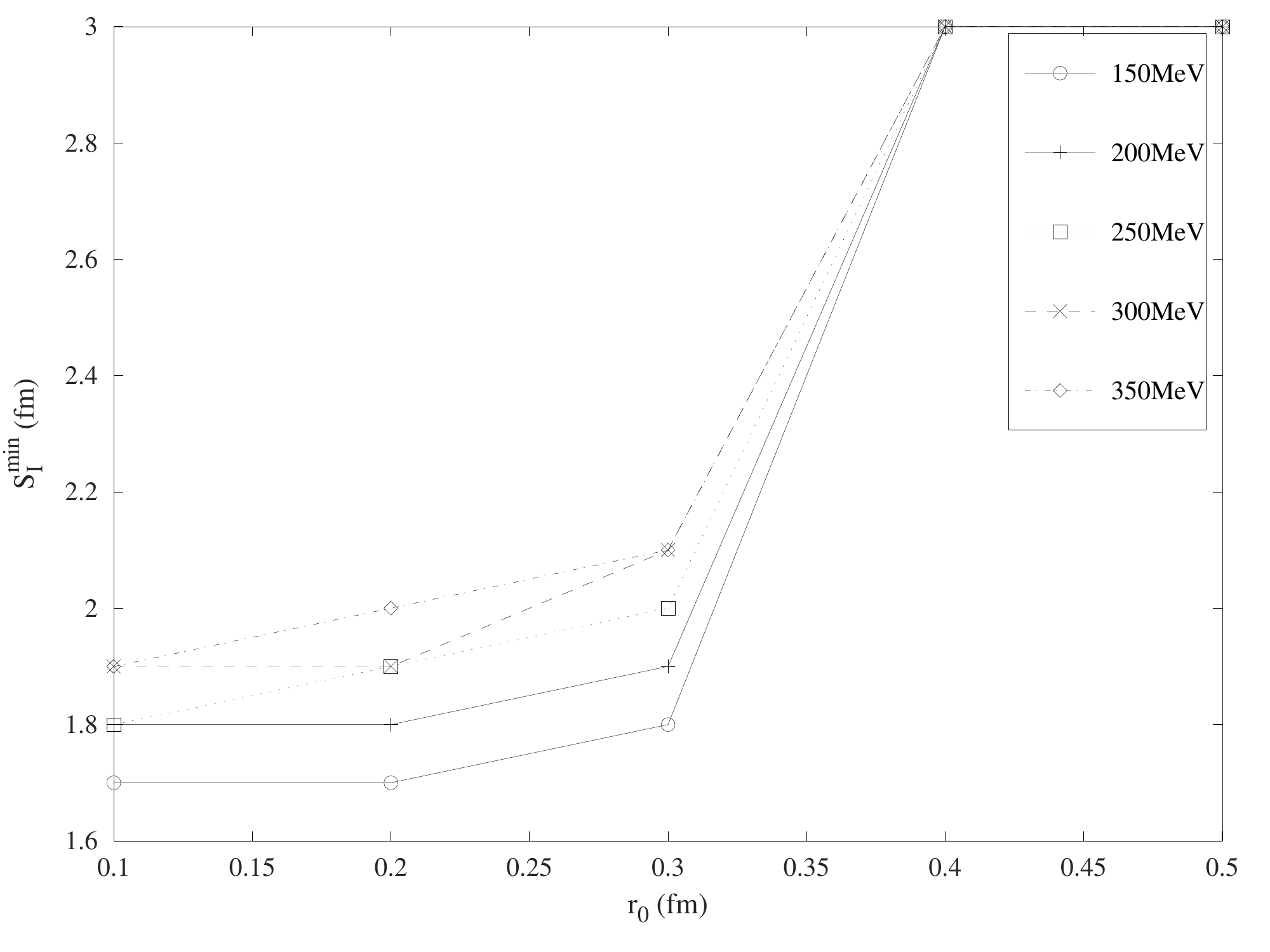}
	\end{subfigure}
	\centering
	\begin{subfigure}[b]{0.3\textwidth}
		\includegraphics[scale=0.2]{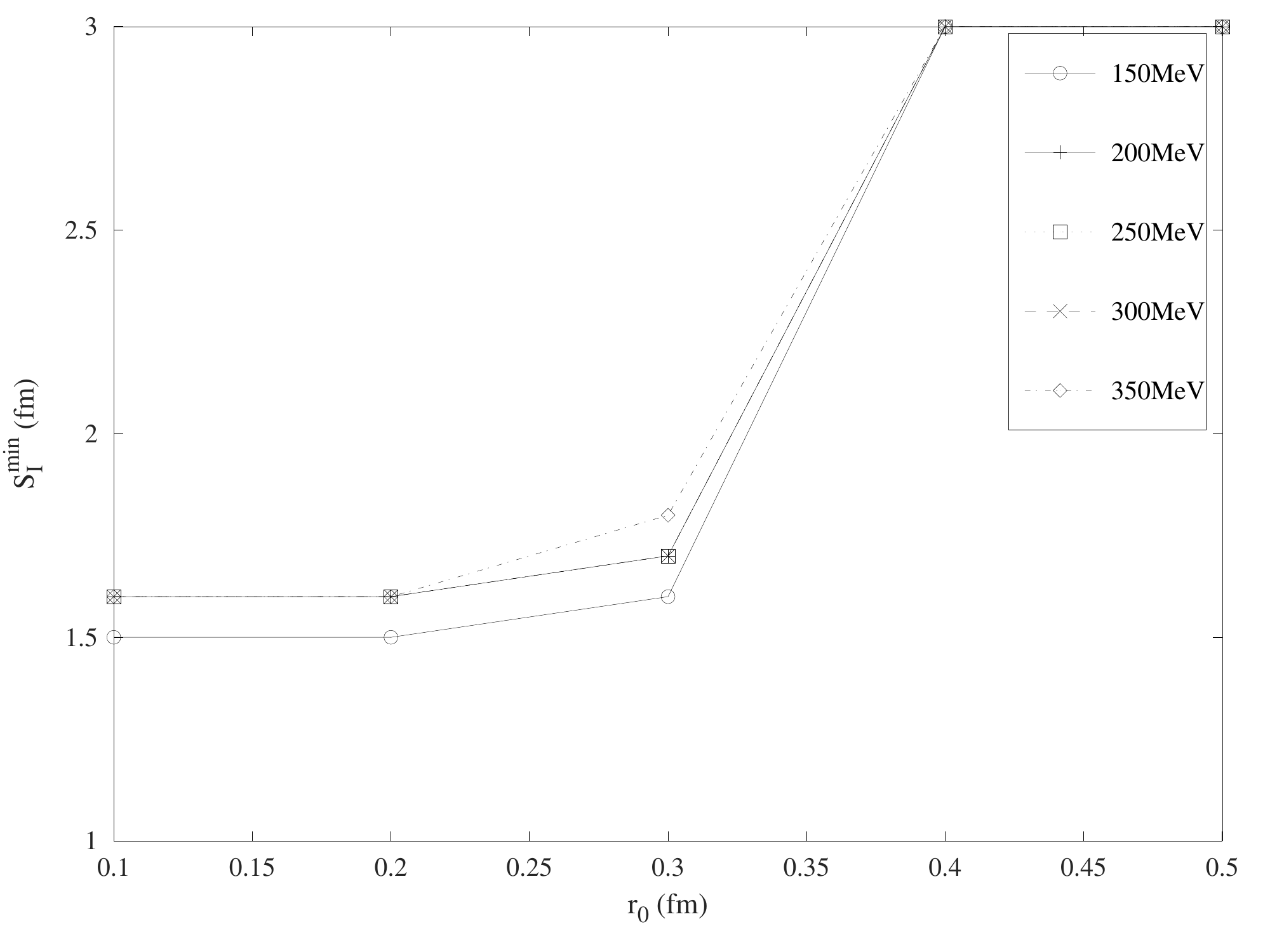}
	\end{subfigure}
	\centering
	\begin{subfigure}[b]{0.3\textwidth}
		\includegraphics[scale=0.2]{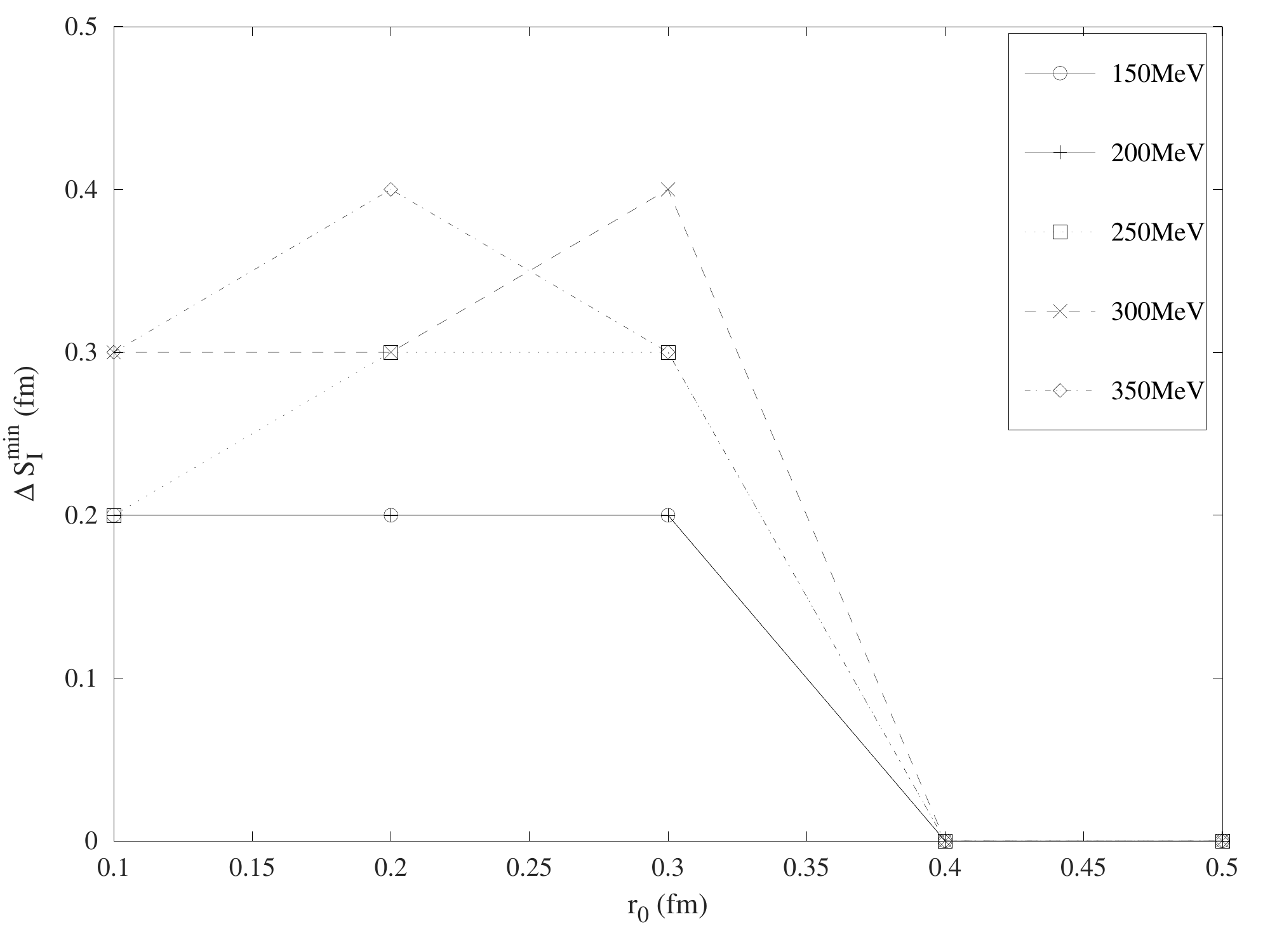}
	\end{subfigure}
	\caption{Variation of the internucleon separation at the minimum of the NN potential ($S_I^{min}$) in the $^1S_0$ (left) and $^3S_1$(center) channels and the difference between the minima of the two channels (right) as functions of $r_0$.}\label{si}
\end{figure}

The strength of repulsion (as measured at $S_I = 0.1\text{ fm}$) decreases as $r_0$ increases and becomes feeble for $r_0\sim b$. The internucleon separation at which the NN potential is minimum ($S_I^{min}$) is also dependent on the value of $r_0$. $S_I^{min}$ increases as $r_0$ decreases (fig. (\ref{si})). \par

\begin{figure}[H]
	\centering
	\begin{subfigure}[b]{0.3\textwidth}
		\includegraphics[scale=0.2]{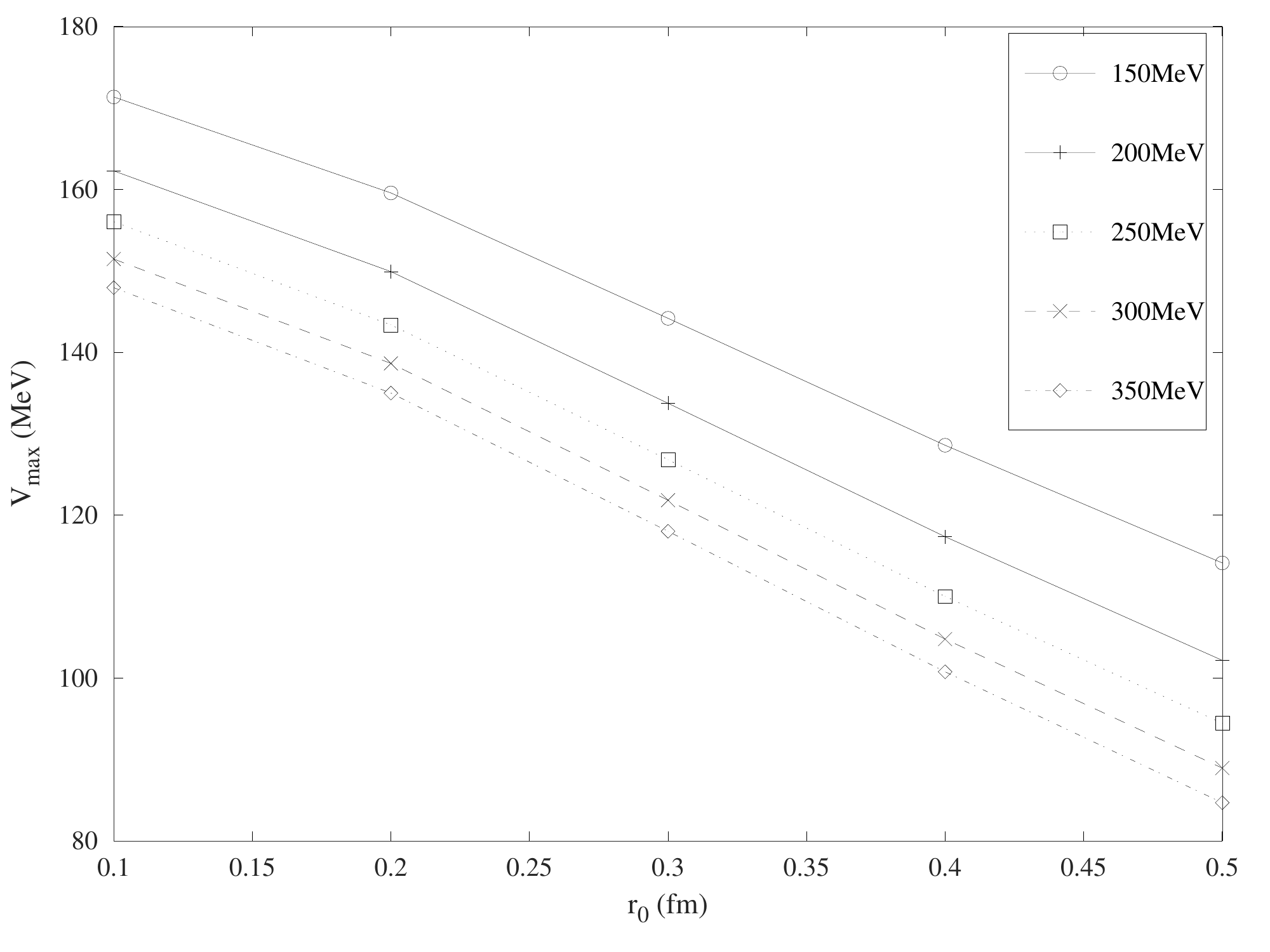}
	\end{subfigure}
	\centering
	\begin{subfigure}[b]{0.3\textwidth}
		\includegraphics[scale=0.2]{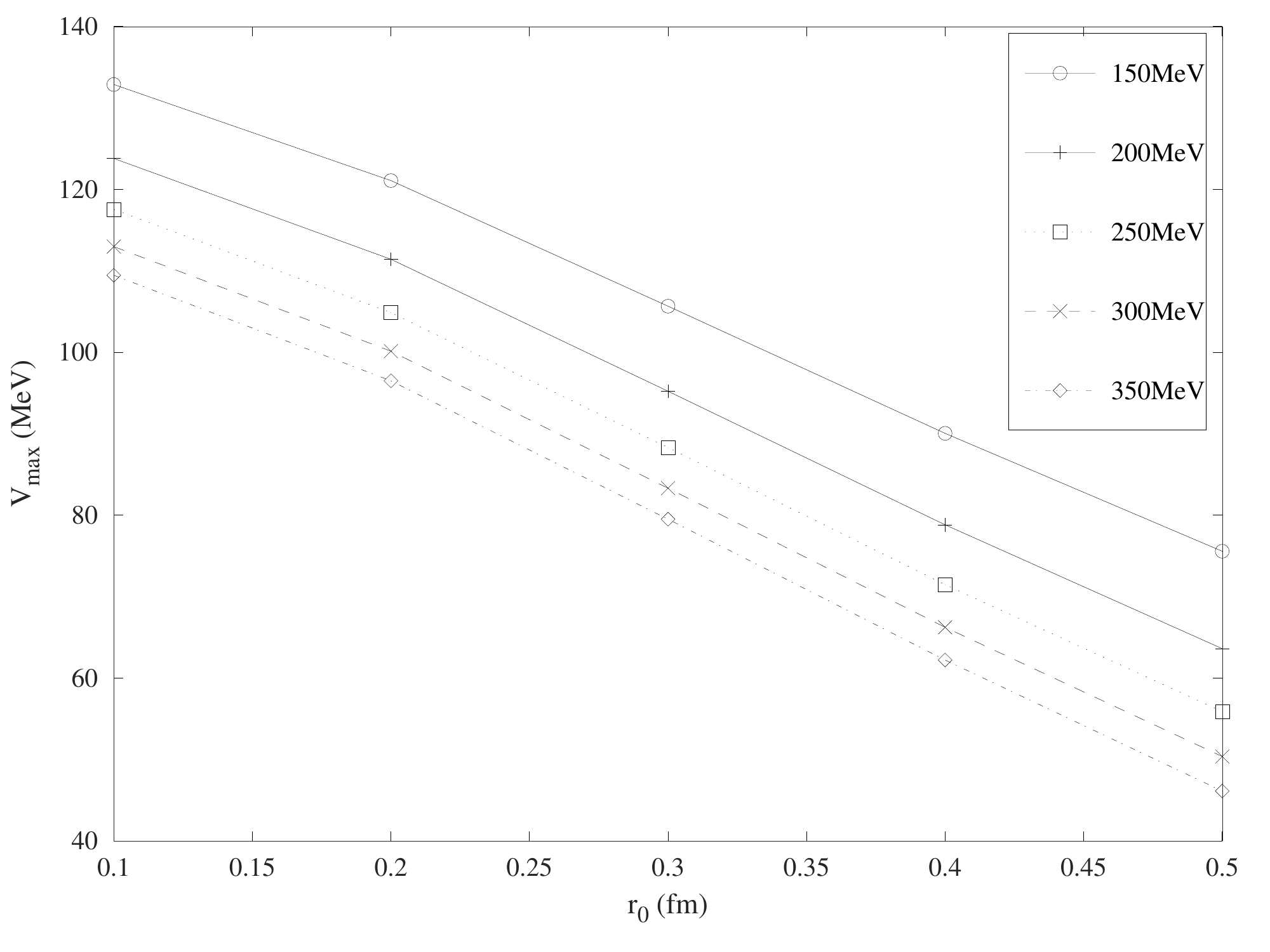}
	\end{subfigure}
	\centering
	\begin{subfigure}[b]{0.3\textwidth}
		\includegraphics[scale=0.2]{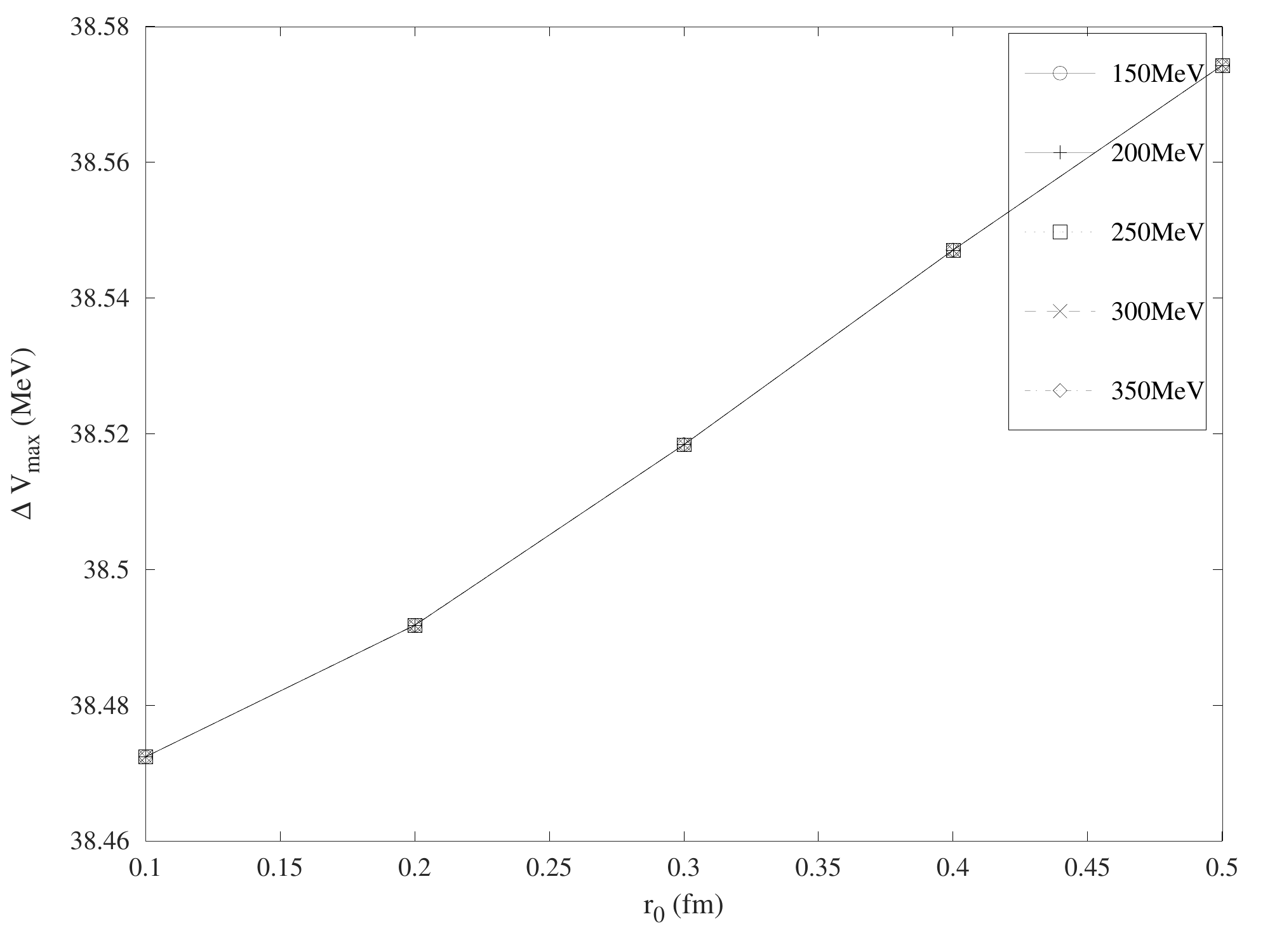}
	\end{subfigure}
	\caption{Variation of the maximum of the NN potential in the $^1S_0$ (left) and $^3S_1$(center) channel and the difference between the maxima of the two two channels (right) as functions of $r_0$.}\label{vmax}
\end{figure}

\subsection{Variation of the mass of the constituent quark}
\begin{figure}[H]
	\centering
	\begin{subfigure}[b]{0.3\textwidth}
		\includegraphics[scale=0.2]{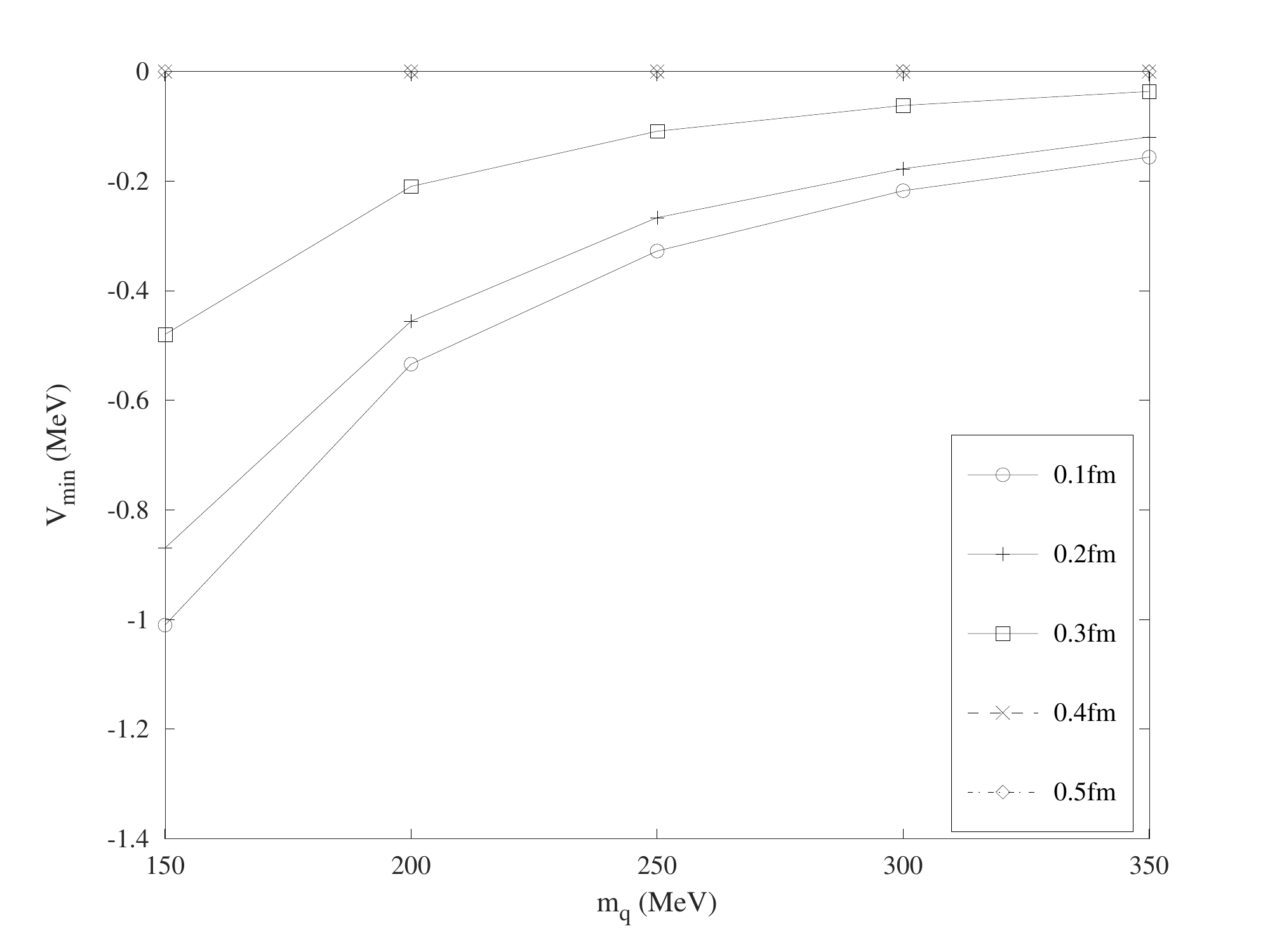}
	\end{subfigure}
	\centering
	\begin{subfigure}[b]{0.3\textwidth}
		\includegraphics[scale=0.2]{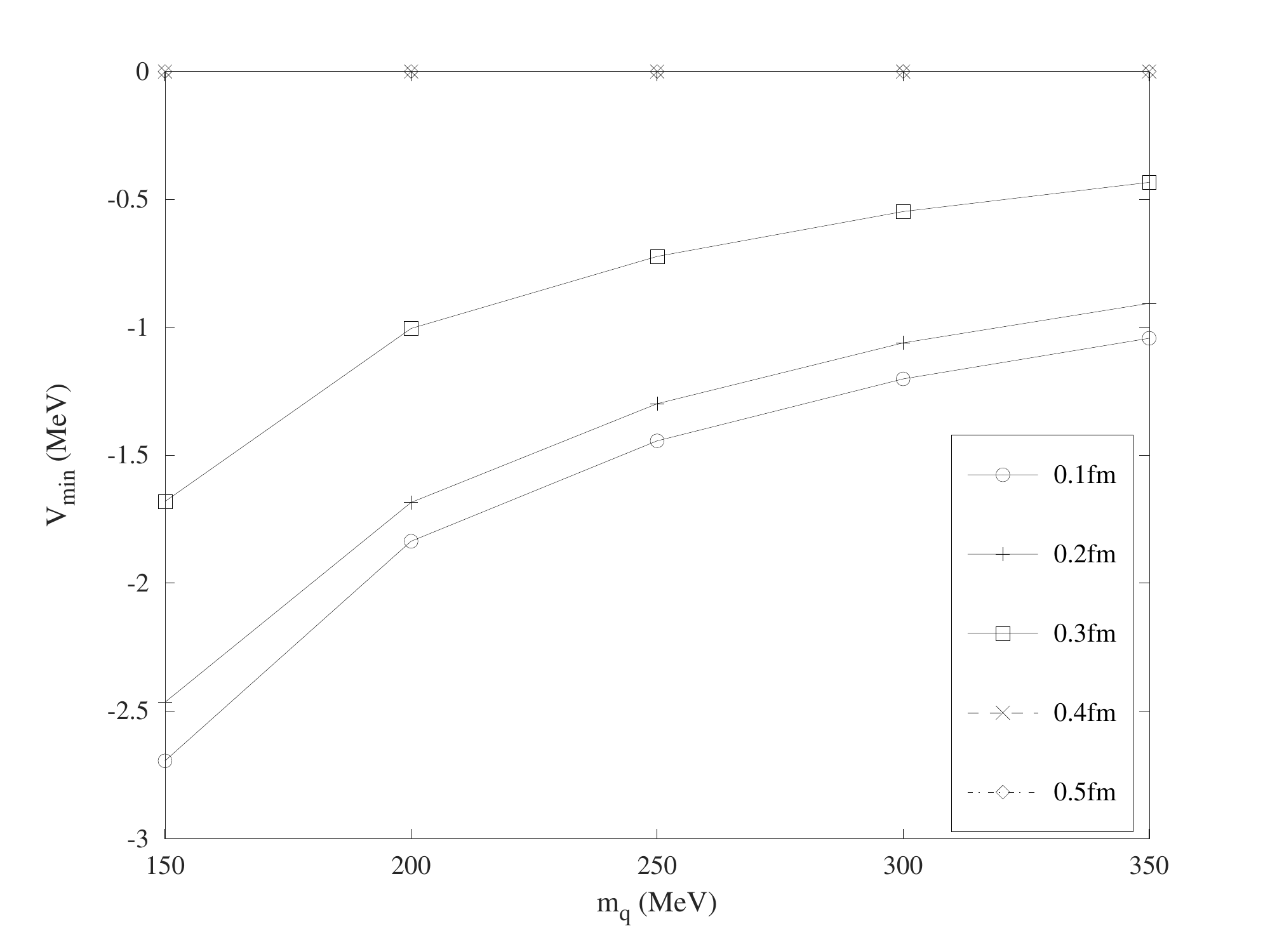}
	\end{subfigure}
	\centering
	\begin{subfigure}[b]{0.3\textwidth}
		\includegraphics[scale=0.2]{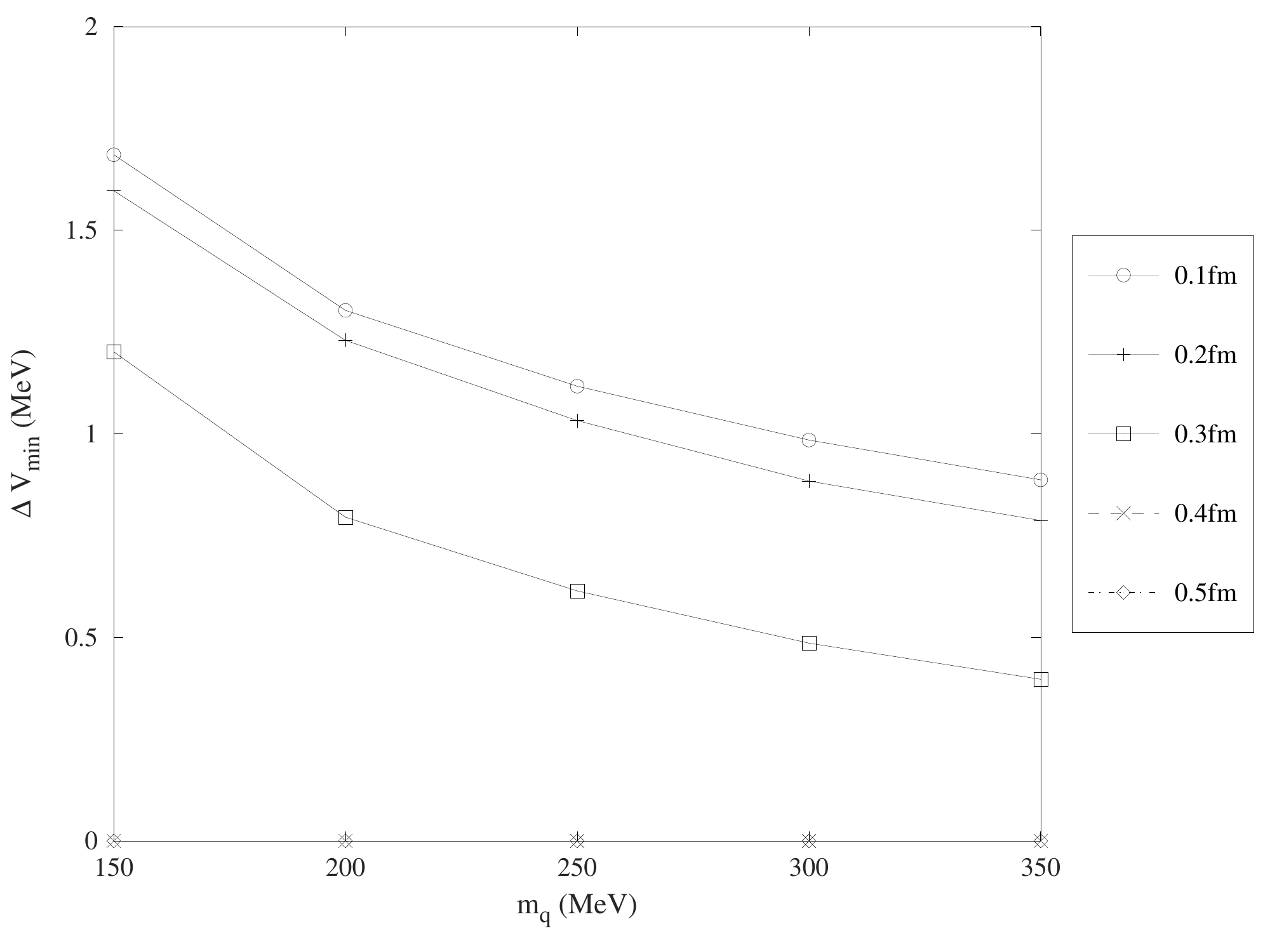}
	\end{subfigure}
	\caption{Variation of the minimum of the NN potential in the $^1S_0$ (left) and $^3S_1$(center) channel and the difference between the maxima of the two two channels (right) as functions of $m_q$.}\label{mvmin}
\end{figure}

The effect of the mass of the constituent quarks ($m_q$) on NN interaction was studied by varying $m_q$ from 150 MeV to 350 MeV. As the mass increases, the magnitude of intermediate range attraction reduces (fig. (\ref{mvmin})). The components of the Hamiltonian sensitive to the variations in the mass are the kinetic energy ($\sim \frac{1}{m_q}$), coulomb part of the OGEP ($\sim m_q^2$) and the confinement energy ($\sim \frac{1}{m_q}$). Thus, the mass dependence of the interaction potential can be given by,
	\begin{equation}\label{eq13}
		V_{NN} = \frac{A_1}{m_q}+A_2 m_q^2+A_3
	\end{equation}
where $A_i$'s are coefficients that depend on the internucleon separation ($S_I$) and $r_0$. The coefficient $A_1$ includes contributions from the kinetic energy term and the confinement term, $A_2$ from the OGEP and $A_3$ from the rest of the Hamiltonian. \par
\begin{figure}[H]
		\centering
	\begin{subfigure}[b]{0.3\textwidth}
		\includegraphics[scale=0.2]{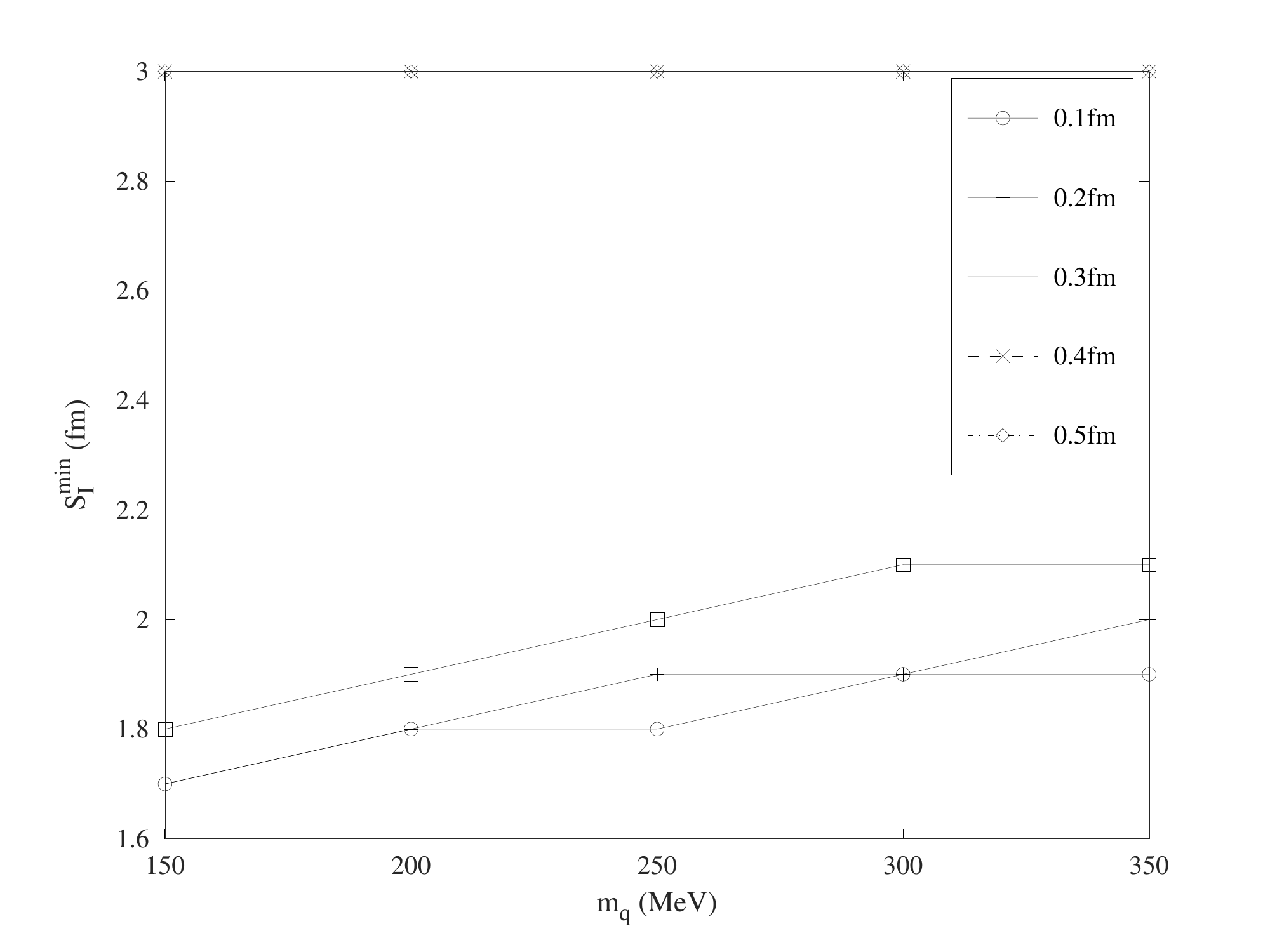}
	\end{subfigure}
	\centering
	\begin{subfigure}[b]{0.3\textwidth}
		\includegraphics[scale=0.2]{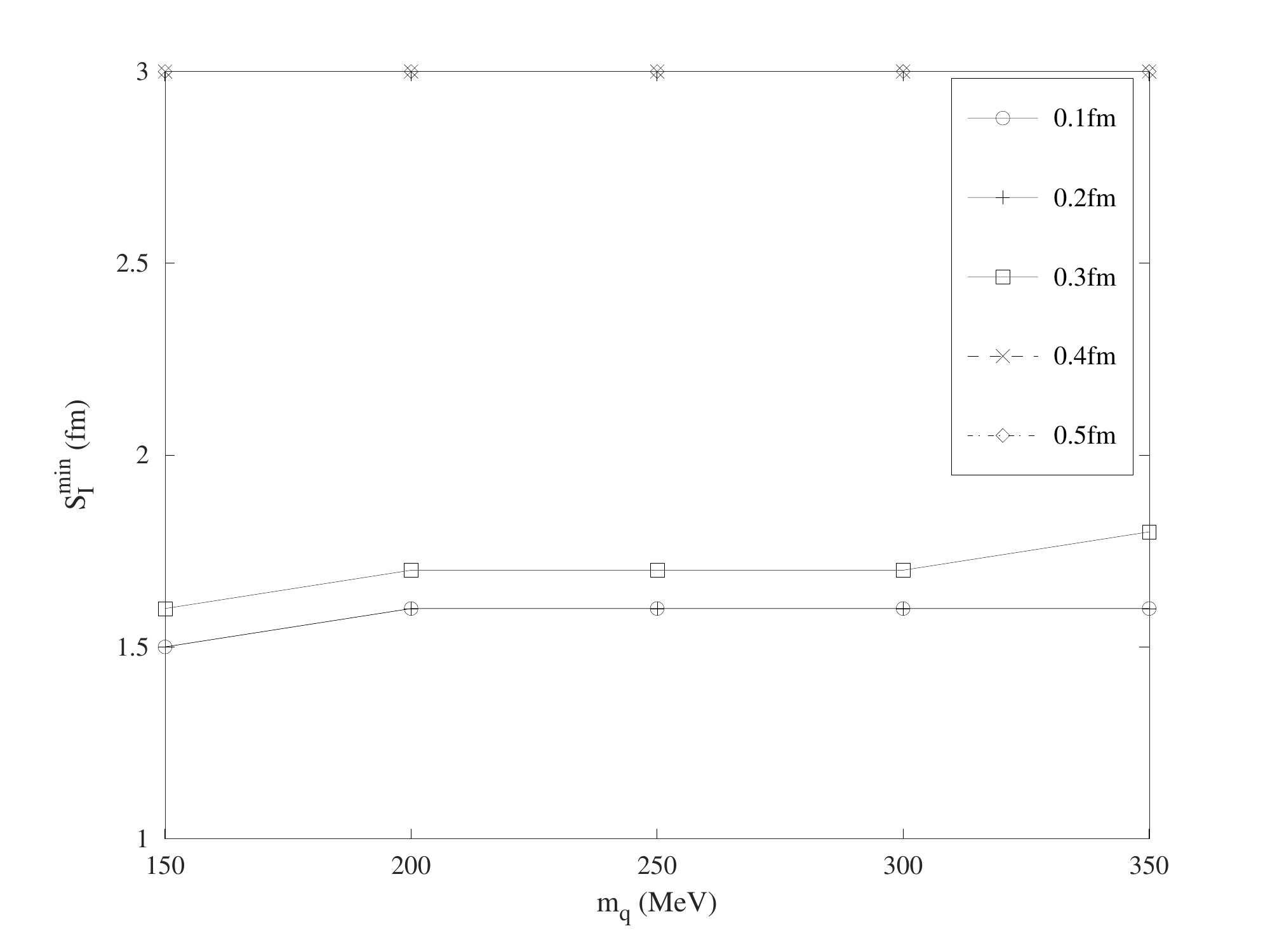}
	\end{subfigure}
	\centering
	\begin{subfigure}[b]{0.3\textwidth}
		\includegraphics[scale=0.2]{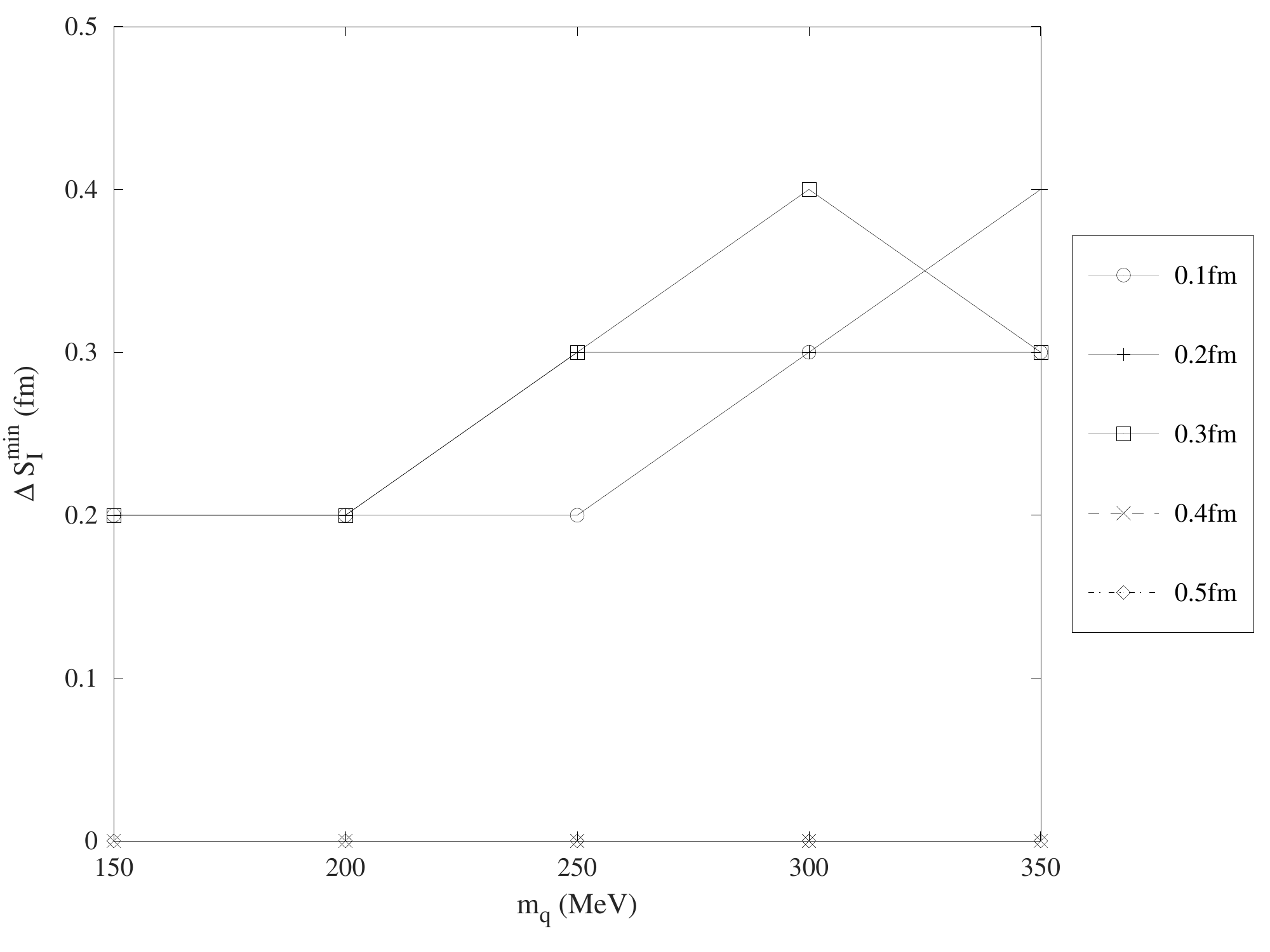}
	\end{subfigure}
	\caption{Variation of the internucleon separation at the minimum of the NN potential ($S_I^{min}$) in the $^1S_0$ (left) and $^3S_1$(center) channels and the difference between the minima of the two channels (right) as functions of $m_q$.}\label{msi}
\end{figure}

Since $\alpha_s\sim m_q^2$, $\frac{\alpha_s}{m_q^2}$ is constant in $m_q$ and the color electric and the color magnetic terms of the OGEP do not vary with $m_q$, the spin dependence of the NN potential appears to be captured entirely in the coefficient $A_3$. This suggests that the difference between the minima of the singlet and the triplet NN potential does not vary with $m_q$. But, fig. (\ref{mvmin}) suggest otherwise. This is because the minima occur at different internucleon separations (fig. (\ref{msi})) and hence the values of $A_i$'s are different for the two channels. Also, since the coulomb term has only a negligible contribution to the NN potential, $A_1$ and $A_3$ form the dominant part of the eq. (\ref{eq13}).

\begin{figure}[H]
	\centering
	\begin{subfigure}[b]{0.3\textwidth}
		\includegraphics[scale=0.2]{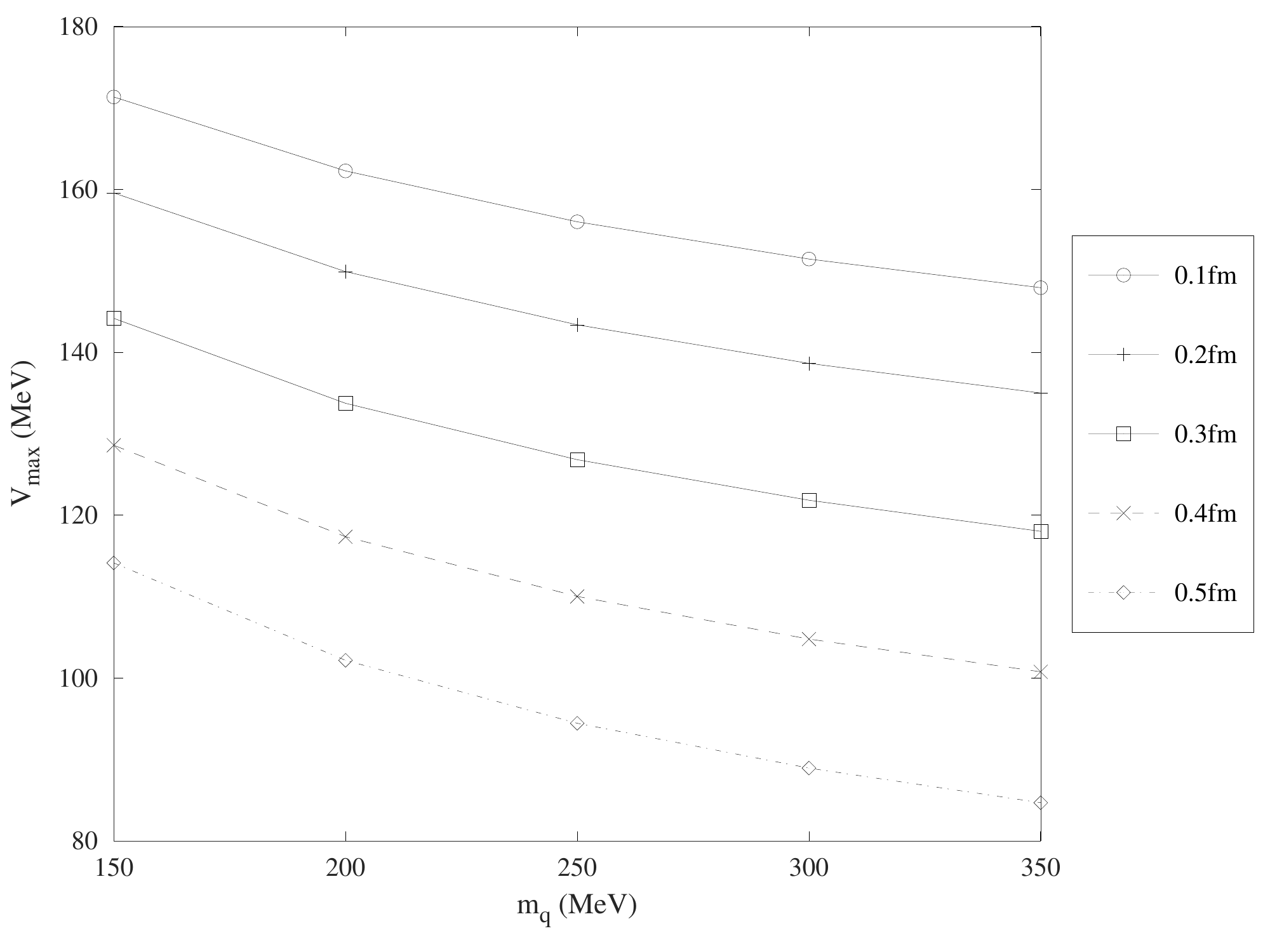}
	\end{subfigure}
	\centering
	\begin{subfigure}[b]{0.3\textwidth}
		\includegraphics[scale=0.2]{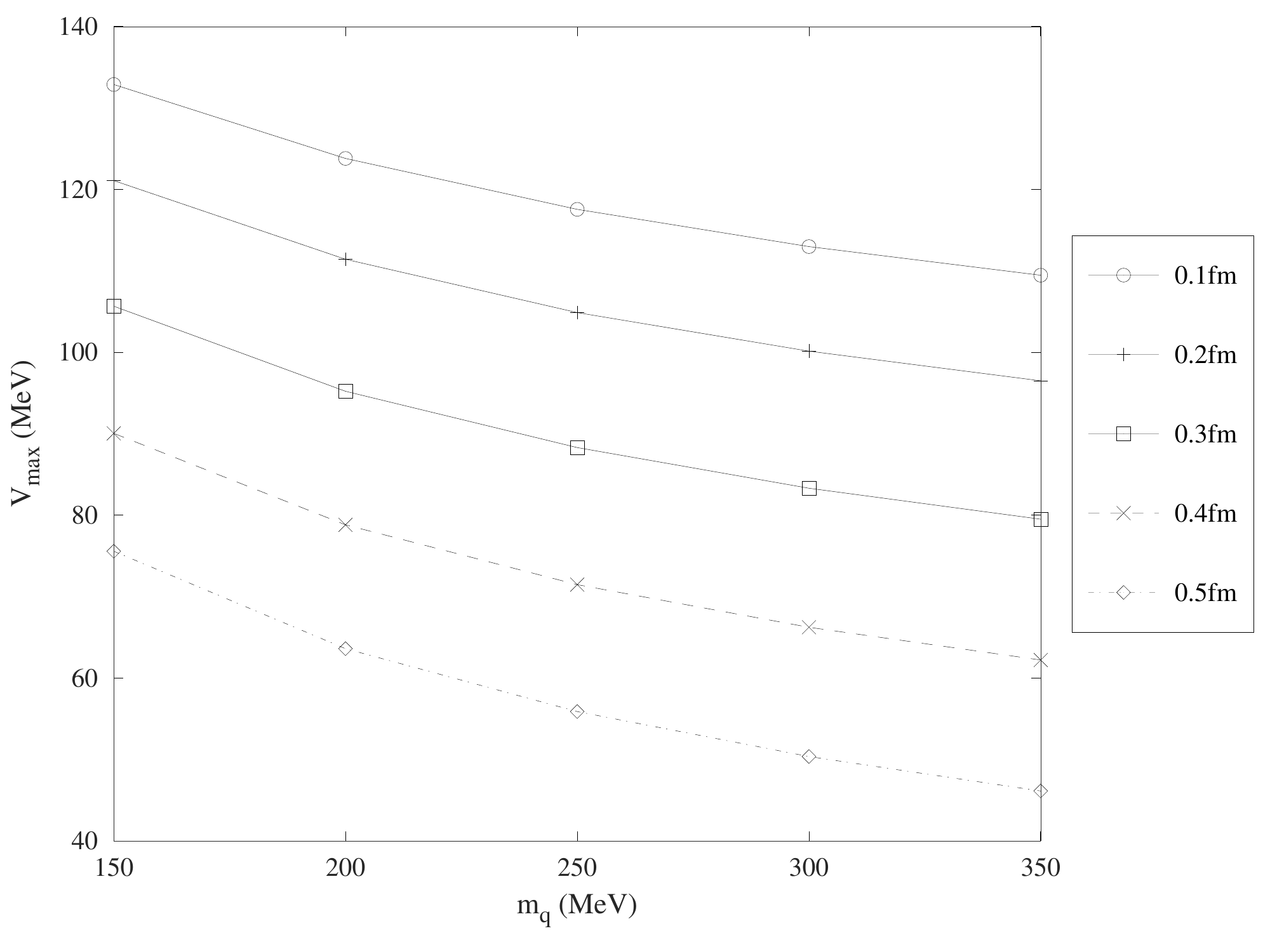}
	\end{subfigure}
	\centering
	\begin{subfigure}[b]{0.3\textwidth}
		\includegraphics[scale=0.2]{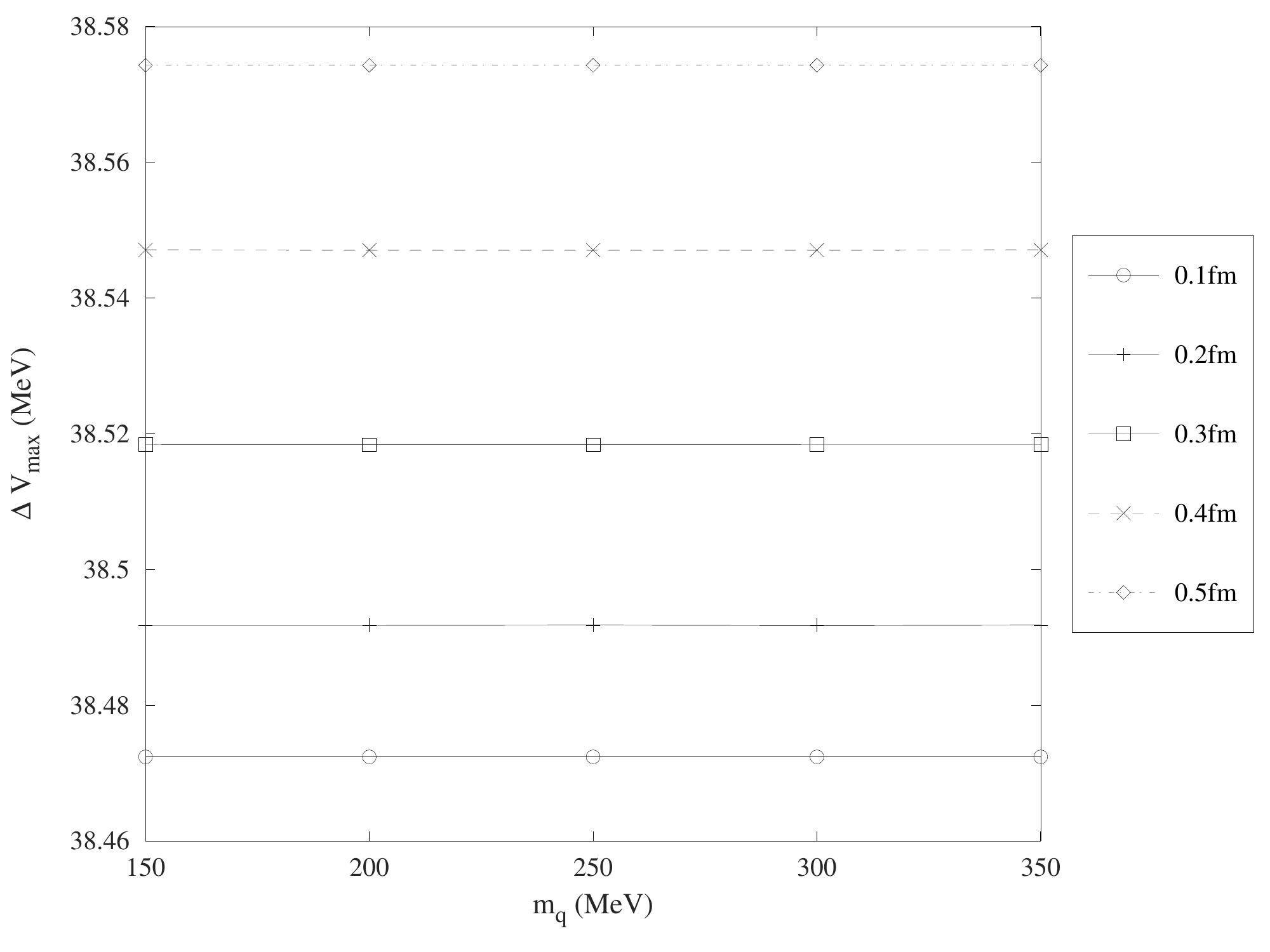}
	\end{subfigure}
	\caption{Variation of the maximum of the NN potential in the $^1S_0$ (left) and $^3S_1$(center) channel and the difference between the maxima of the two two channels (right) as functions of $m_q$.}\label{mvmax}
\end{figure}

The short range repulsion is also affected by the mass of the constituent quarks as shown in fig. (\ref{mvmax}). The magnitude of repulsion reduces as the constituent mass increases again ascertaining that the terms involving the inverse power of mass ($\frac{1}{m_q}$) are responsible for the variation of the NN potential with the mass of the constituent quarks. However, the variation of the spin dependent terms could not be studied as the maxima occur at $S_I\sim 0$ and the model breaks down at such short distances. In the range of $S_I$ studied, the difference between the strength of repulsion in the two channels does not vary with $m_q$.

\begin{figure}[H]
	\centering
	\begin{subfigure}[b]{0.45\textwidth}
		\includegraphics[scale=0.35]{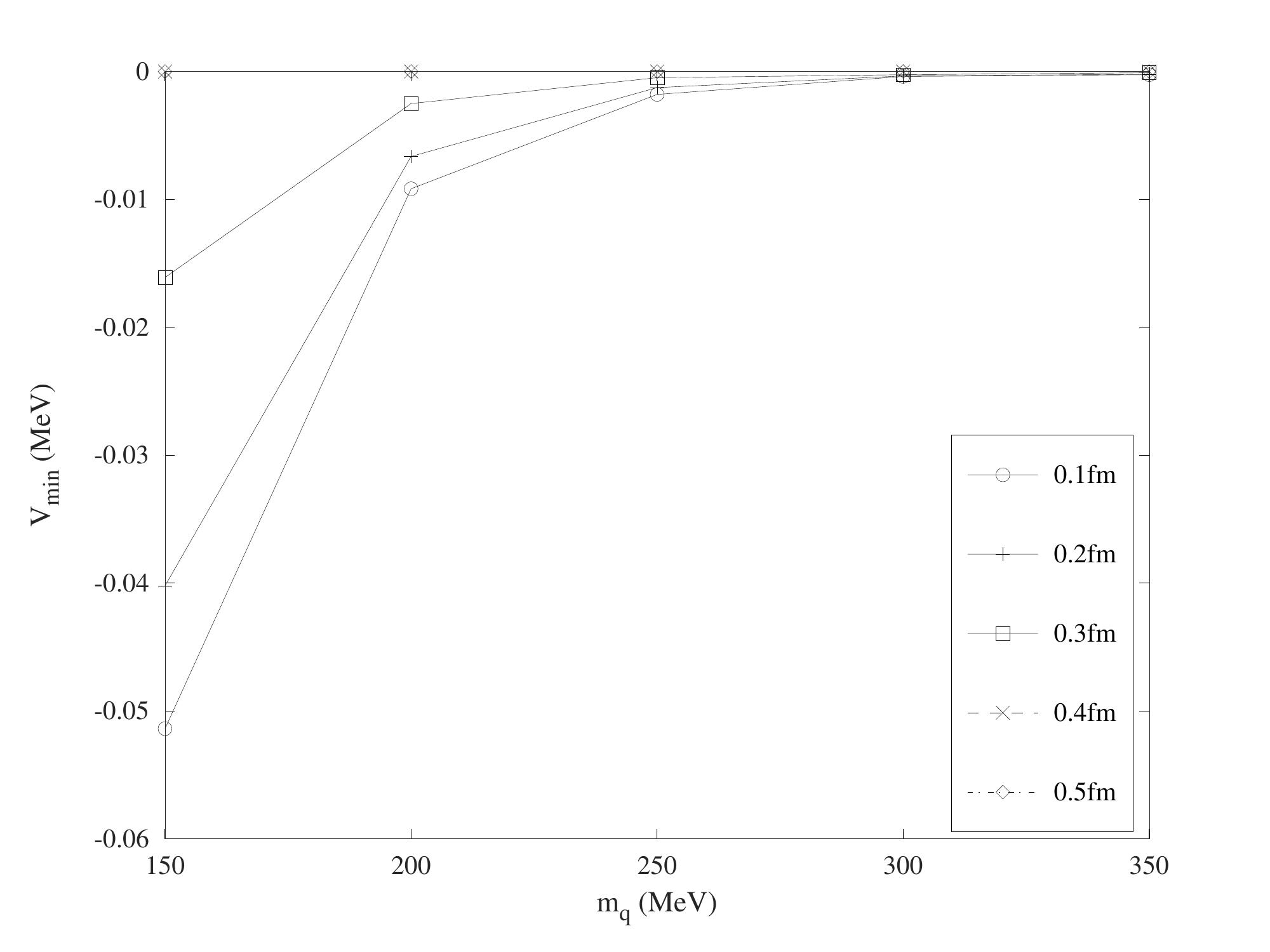}
	\end{subfigure}
	\centering
	\begin{subfigure}[b]{0.45\textwidth}
		\includegraphics[scale=0.35]{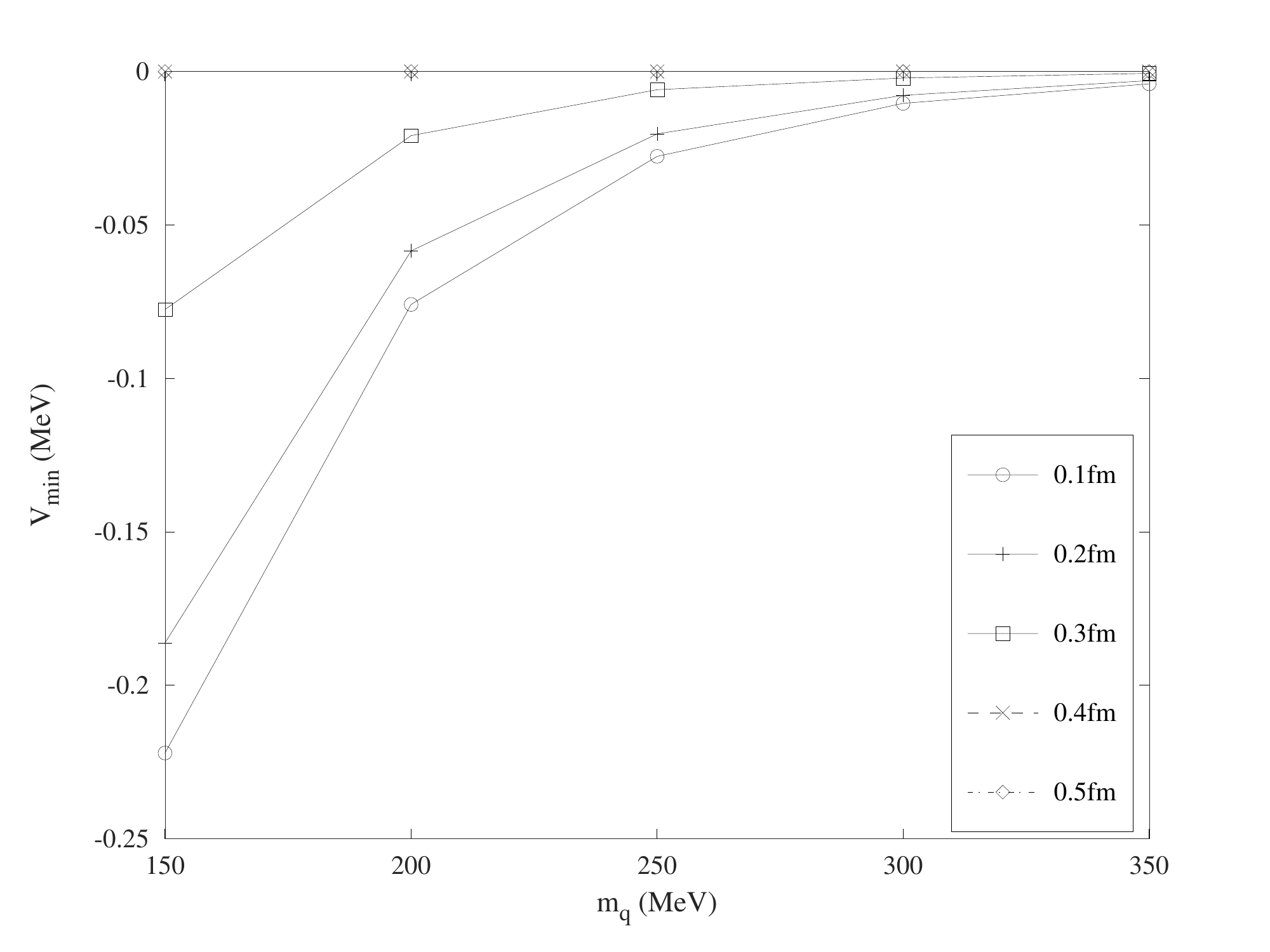}
	\end{subfigure}
	\caption{Variation of the minimum of the NN potential in the $^1S_0$ (left) and $^3S_1$(right) channel and the as functions of $m_q$ in the absence of III.}\label{nnoi}
\end{figure}

\begin{figure}[H]
	\centering
	\begin{subfigure}[b]{0.45\textwidth}
		\includegraphics[scale=0.35]{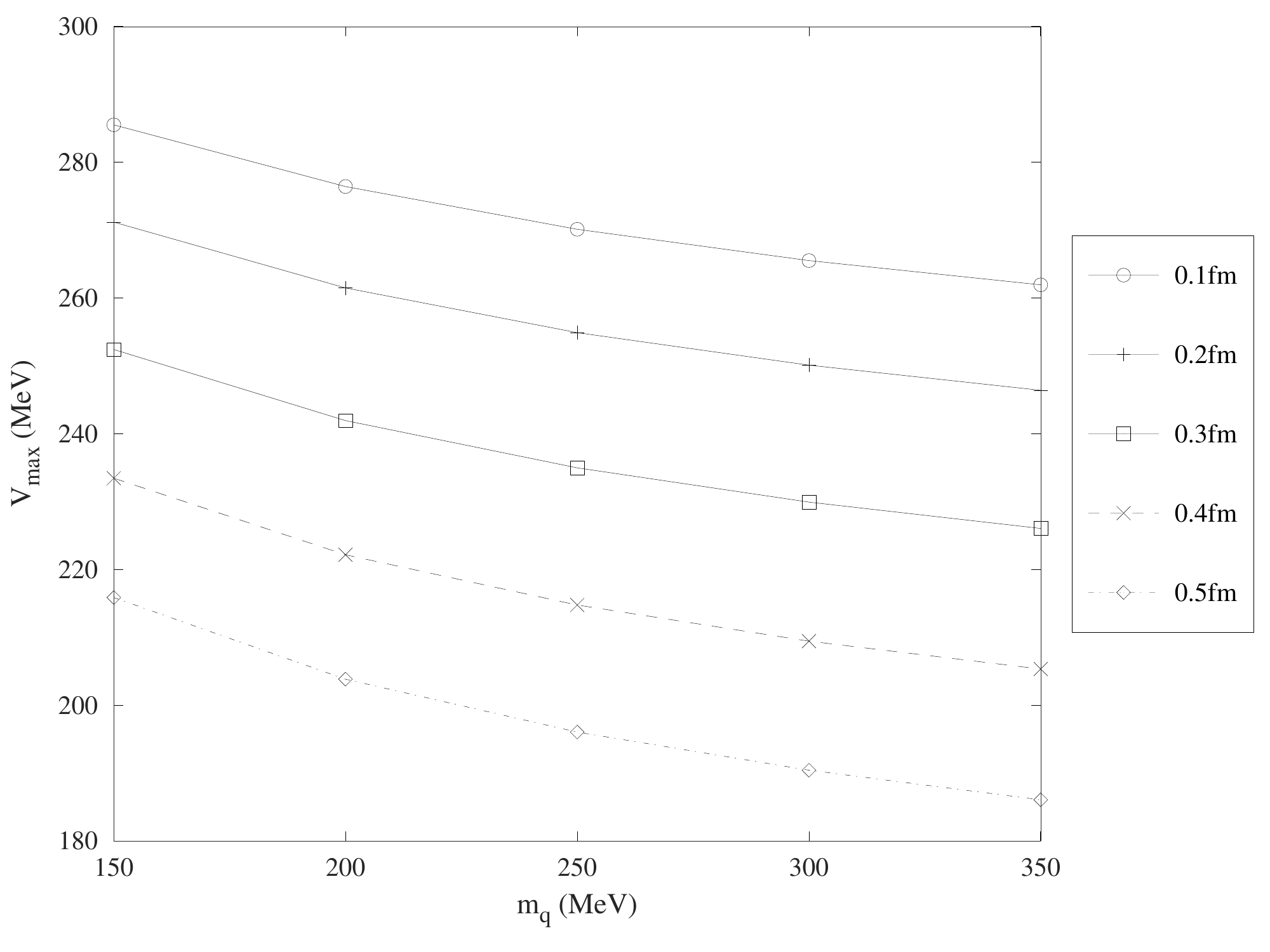}
	\end{subfigure}
	\centering
	\begin{subfigure}[b]{0.45\textwidth}
		\includegraphics[scale=0.35]{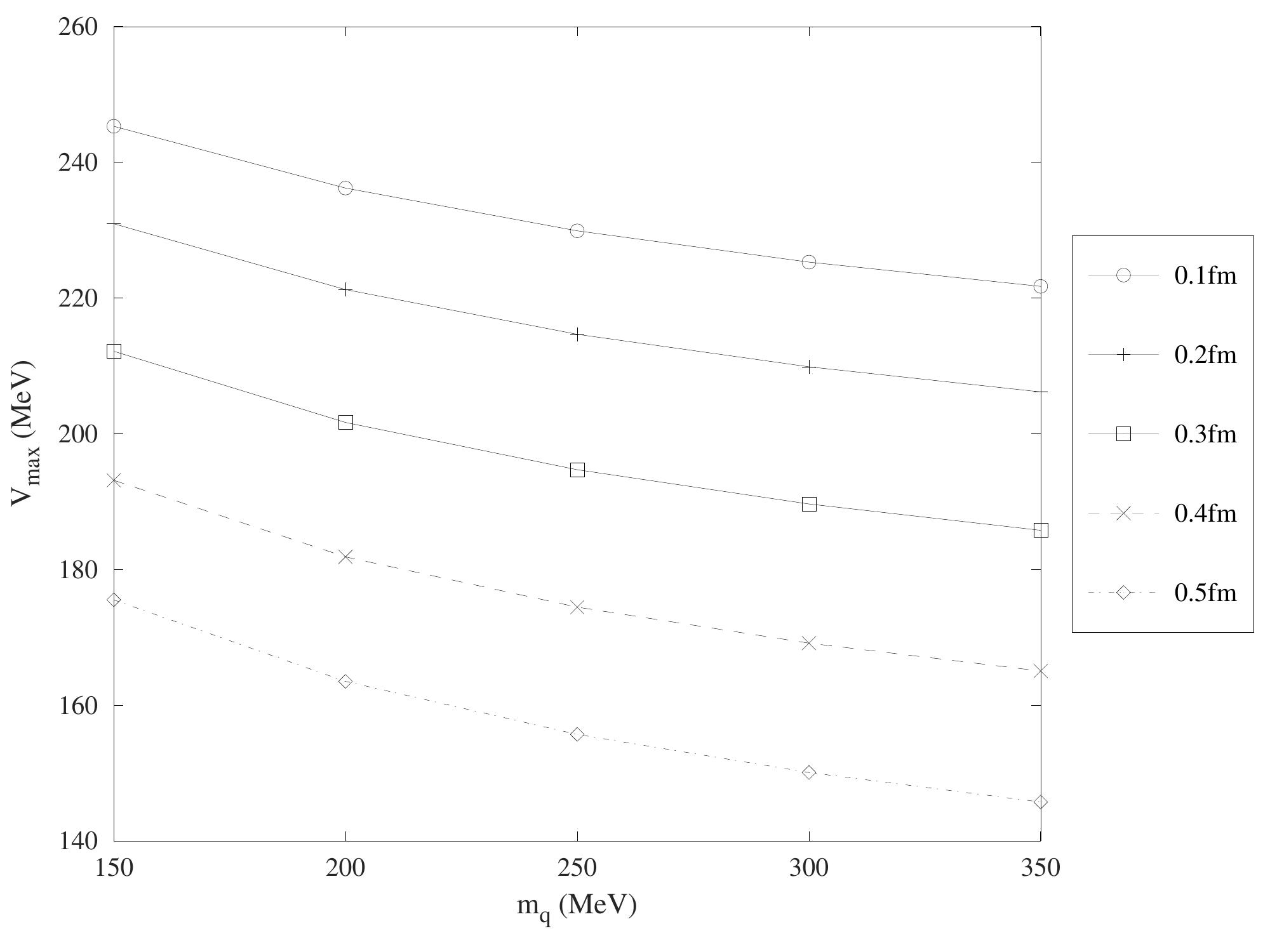}
	\end{subfigure}
	\caption{Variation of the maximum of the NN potential in the $^1S_0$ (left) and $^3S_1$(right) channel as functions of $m_q$ in the absence of III.}\label{xnoi}
\end{figure}

In the absence of the III, the intermediate range interaction in feeble ($<0.25$ MeV) (fig. (\ref{nnoi})), whereas the short-range repulsion more than doubles in magnitude (fig. (\ref{xnoi})). Thus, it appears that even though it is the chiral symmetry breaking and hence the III that provides the dynamical mass and the finite size for the quarks, III cannot be just be discarded from NN interaction models. However, deeper study is needed to ascertain the true nature of these interactions.

\begin{figure}[H]
		\centering
	\begin{subfigure}[b]{0.3\textwidth}
		\includegraphics[scale=0.2]{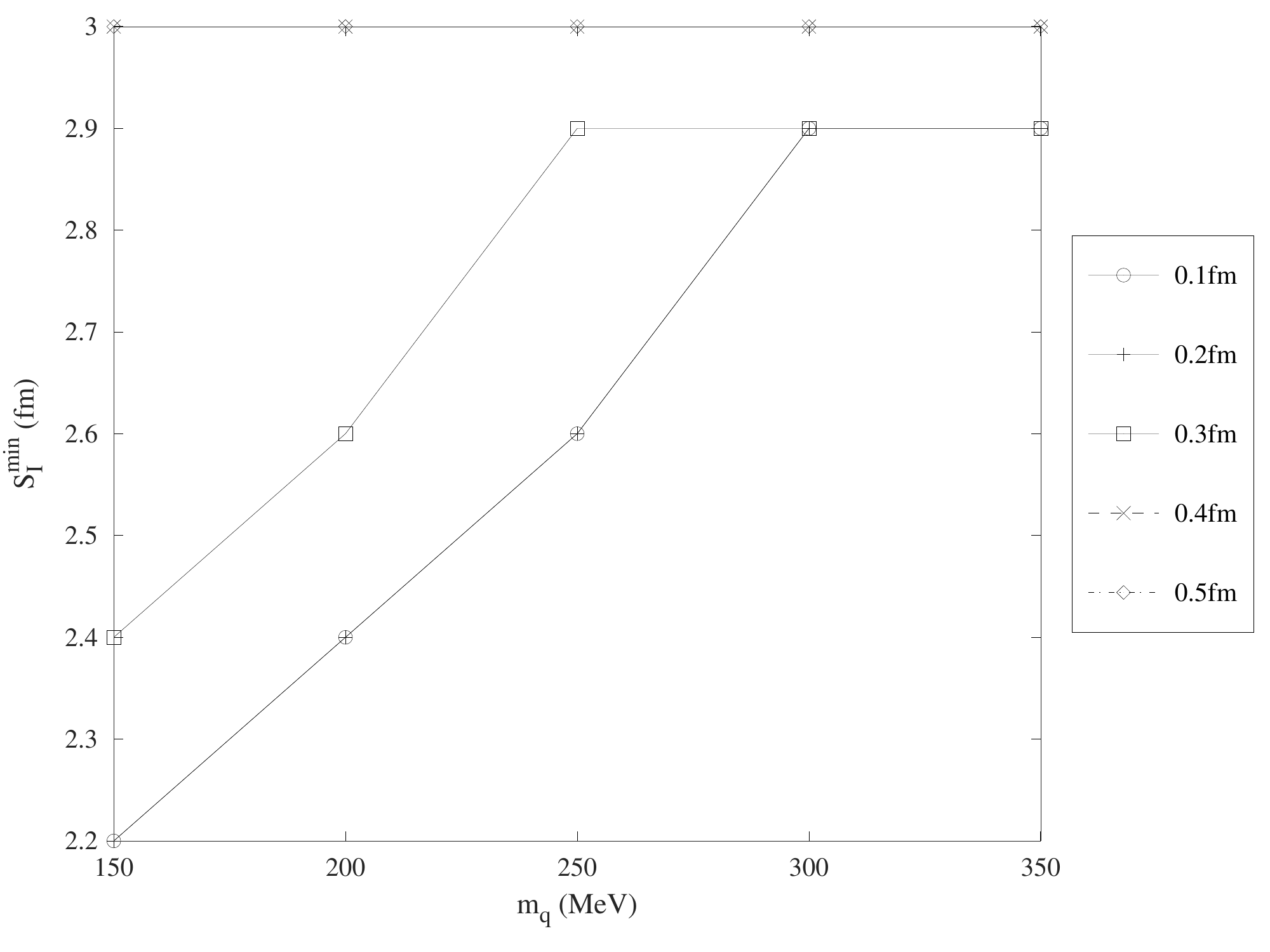}
	\end{subfigure}
	\centering
	\begin{subfigure}[b]{0.3\textwidth}
		\includegraphics[scale=0.2]{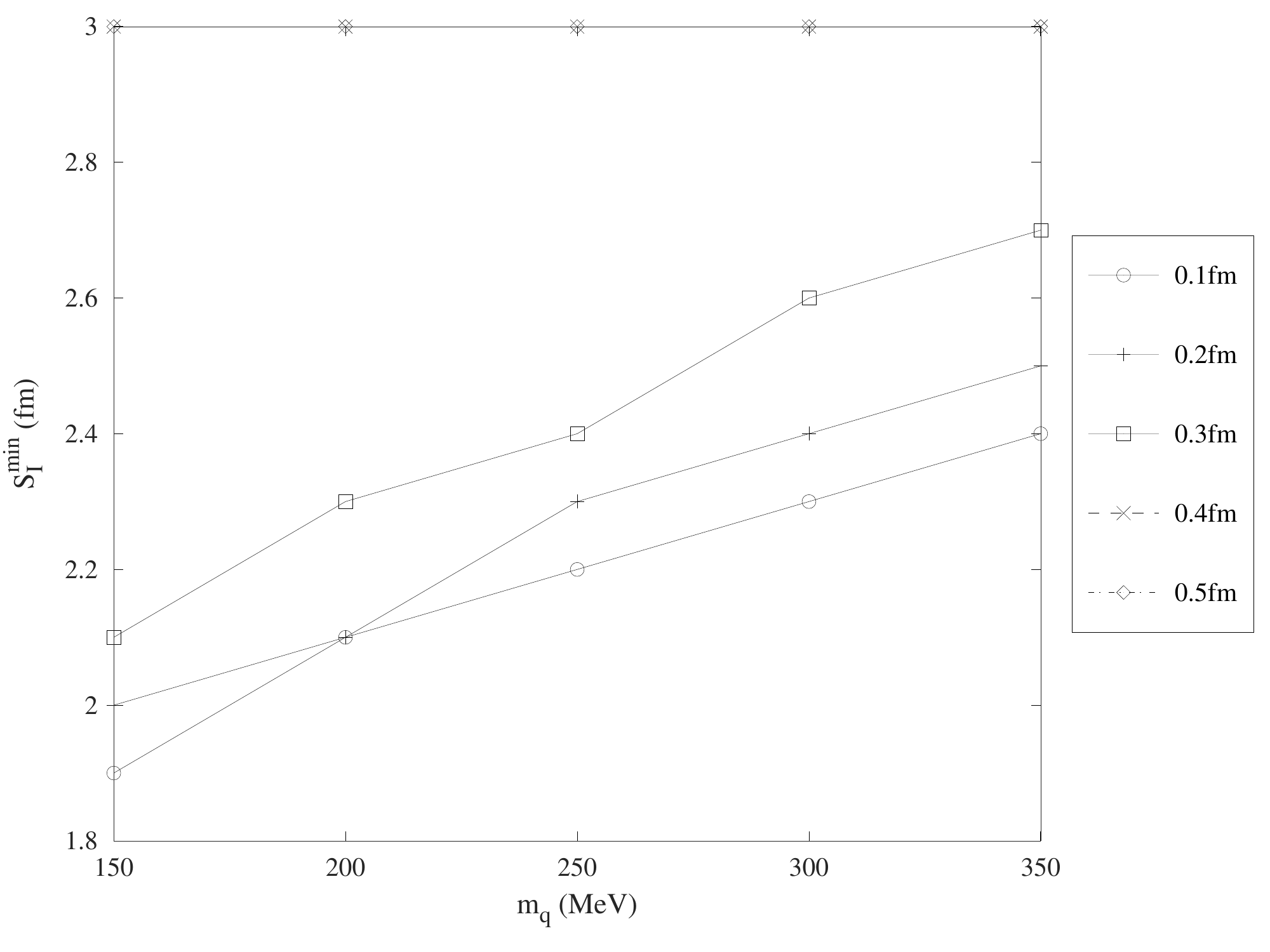}
	\end{subfigure}
	\centering
	\begin{subfigure}[b]{0.3\textwidth}
		\includegraphics[scale=0.2]{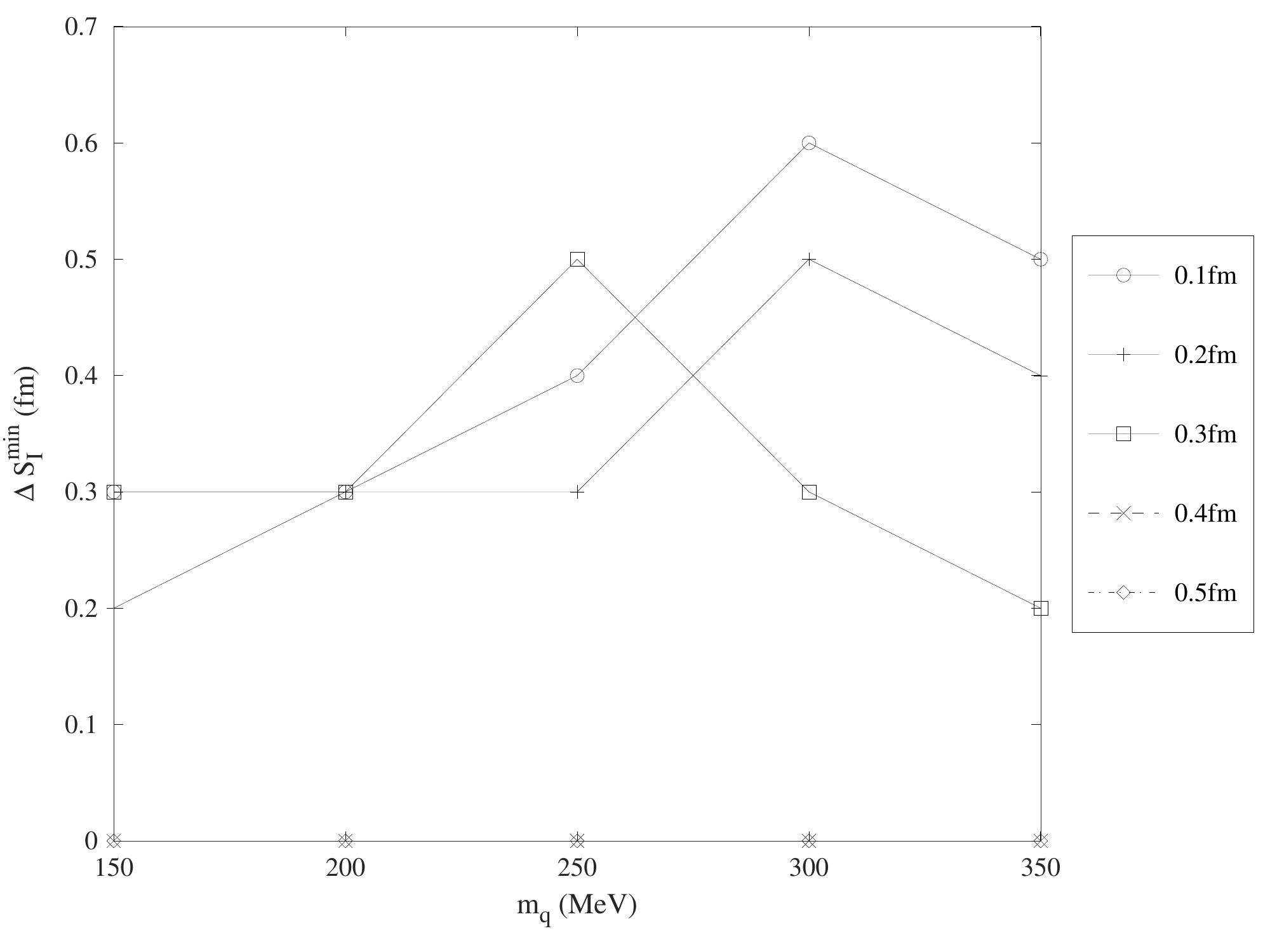}
	\end{subfigure}
	\caption{Variation of the internucleon separation at the minimum of the NN potential ($S_I^{min}$) in the $^1S_0$ (left) and $^3S_1$(center) channels and the difference between the minima of the two channels (right) as functions of $m_q$ in the absence of III.}\label{sinoi}
\end{figure}

Also, the variations in $S_I^{min}$ as a function of $m_q$ is large (fig. (\ref{sinoi})). The III, thus, provides stability for the system against the repulsive forces present due to the exchange of gluons.

\section{Summary and Outlook}
In this work, we have employed the NRQM in the frame work of RGM by varying the $r_0$ from 0.1 fm to 0.6 fm to study the effects of the finite size of the quark core and its quark mass on NN interaction. From our analysis, as the quark mass decreases the short range repulsion increases and simultaneously the attraction in the intermediate range also decreases gradually. The magnitude and the range of the short range repulsion decreases as $r_0$ increases and the depth of the attractive well is maximum when $r_0\sim b$. Hence, the attractive part of the NN interaction and the internucleon separation corresponding to minimum of the potential depend on quark size parameter. The total short range repulsion of the NN potential also reduces as the quark size parameter increases. The $\alpha_s$ and $W$ depend on the size of the constituent quarks and $\alpha_s$ becomes greater than 1 for large values of $r_0$. We also observe that the III cannot be replaced by the inclusion of quark size parameter as the instantons play a fundamental role in strong interactions. \par
Even though the behavior of the NN potential under the influence of smearing potential has been studied, the true influence of the smearing potential can be evaluated only by studying the phase shifts and the cross section for NN scattering. Deeper study is needed to ascertain the relation between $r_0$, $m_q$ and the III.

\bibliographystyle{unsrtnt}
\bibliography{NN_Finite}
\end{document}